\documentclass[12pt,a4paper]{article}
\usepackage[utf8]{inputenc}
\usepackage[T1]{fontenc}
\usepackage{amsmath,hyperref,bbm,cite}
\usepackage{amsfonts}
\usepackage{amssymb,comment}
\usepackage{graphicx}
\usepackage{transparent}
\usepackage{notoccite}
\usepackage{ulem}
\usepackage[left=2cm, right=2cm]{geometry}
\usepackage{todonotes}
\usepackage{xcolor}
\usepackage{physics}
\usepackage{bbold}
\usepackage{ntheorem}
\usepackage[overlay,absolute]{textpos}

\newcommand{\be}{\begin{equation}}
\newcommand{\ee}{\end{equation}}
\newcommand{\bea}{\begin{eqnarray}}
\newcommand{\eea}{\end{eqnarray}}

\newcommand\PlaceText[3]{%
\begin{textblock*}{10in}(#1,#2)
#3
\end{textblock*}
}%

\begin{document}
\sloppy

\PlaceText{165mm}{15mm}{11 October 2023}
\begin{center}
{\LARGE\bf Cobordism and Bubbles of Anything}

\vspace{.5cm}
{\LARGE\bf in the String Landscape}

\vspace{.5cm}
\end{center} 

\vspace*{0.8cm}
\thispagestyle{empty}

\centerline{\large 
Bjoern Friedrich$^{1}$,
Arthur Hebecker$^1$, Johannes Walcher$^{1,2}$}
\vspace{0.5cm}
 
\begin{center}
$^1${\it Institute for Theoretical Physics, Heidelberg University,}
\\
{\it Philosophenweg 19, 69120 Heidelberg, Germany}\\[0.5ex]  
\end{center} 
\vspace{-.4cm}
 
\begin{center}
$^2${\it Institute for Mathematics, Heidelberg University,}
\\
{\it Im Neuenheimer Feld 205, 69120 Heidelberg, Germany}\\[0.5ex]  
\end{center} 

\vspace{.25cm}
\centerline{\small\textit{E-Mail:} \href{mailto:friedrich@thphys.uni-heidelberg.de}{friedrich@thphys.uni-heidelberg.de}, \href{mailto:a.hebecker@thphys.uni-heidelberg.de}{a.hebecker@thphys.uni-heidelberg.de},}
\centerline{\href{mailto:walcher@uni-heidelberg.de}{walcher@uni-heidelberg.de}}

\vspace*{.5cm}
\begin{abstract}\normalsize
We study bubble of nothing decays and their reverse processes, the creation of vacua through `bubbles of something', 
in models of the Universe based on string theory.
From the four-dimensional perspective, the corresponding gravitational instantons contain an end-of-the-world (ETW) boundary or brane, realized by the internal manifold shrinking to zero size.
The existence of such ETW branes is predicted by the Cobordism Conjecture. We develop the 4d EFT description of such boundaries at three levels: First, by generalizing the Witten bubble through an additional defect. 
Second, by replacing the compact $S^1$ with a Calabi-Yau orientifold and allowing it to shrink and disappear through a postulated defect. 
Third, we describe an ETW brane construction for type IIB Calabi-Yau orientifold compactifications with O3/O7 planes through an appropriate additional O5 orientifolding.
Our 4d EFT formalism allows us to compute the decay/creation rates for bubbles of anything 
depending on two parameters: The size of the relevant defect and its tension a.k.a. the induced (generalized) deficit angle.

\vspace*{.4cm}
\noindent
\end{abstract}
\vspace{10pt}

\newpage

\tableofcontents 

\newpage

\section{Introduction}
\begin{figure}[h]
	\centering
	\includegraphics[width=0.9\linewidth]{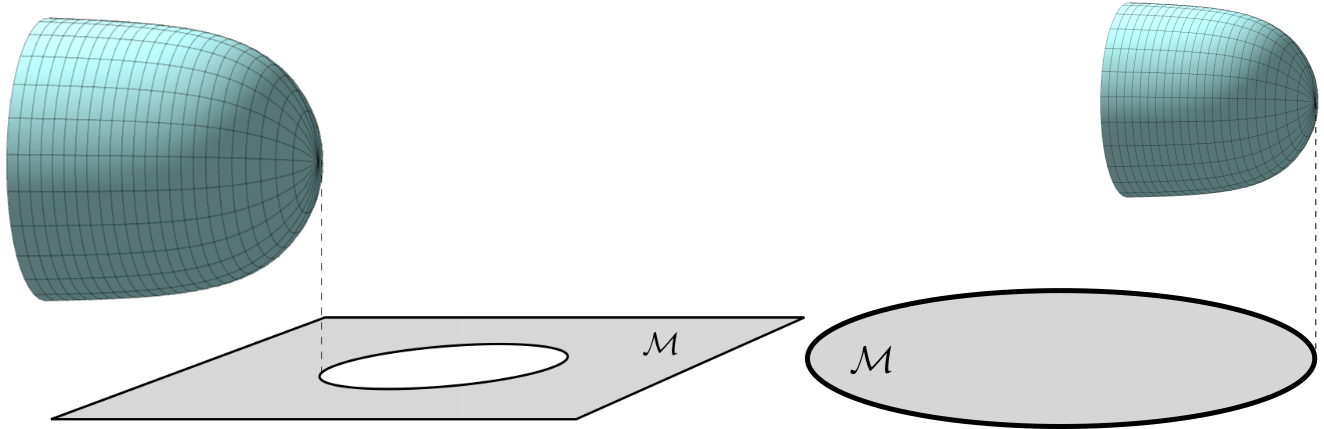}
	\caption{Illustration of the bubble of nothing (left) and the bubble of something (right) instanton. In both cases, the four-dimensional space $\mathcal{M}$ has a spherical boundary which coincides with the locus where the internal manifold, shown above $\mathcal{M}$, shrinks to zero size. 
 }%
	\label{fig_bubble_of_anything}
\end{figure}

In the presence of multiple vacua, for example in Calabi-Yau compactifications of string theory, the decay of one vacuum into another is 
in general a common phenomenon. Many such processes admit an effective semi-classical description involving scalar fields in potentials with multiple critical points \cite{Coleman:1980aw,Hawking:1981fz}.
Witten \cite{Witten:1981gj} discovered that, in addition, for some vacua arising from compactifications of a higher-dimensional theory, 
a decay process to `nothing' also exists, whose effective field theory description is much less obvious.
Specifically in the case of an $S^1$ compactification from 5d to 4d, the relevant instanton is characterized by the $S^1$ shrinking to zero size over the locus of an $S^3$, cf.~l.h.s.~of Fig.~\ref{fig_bubble_of_anything}. The inside of the $S^3$ is empty while far away from that region the geometry approaches the vacuum.
\begin{figure}[h]
	\centering
	\includegraphics[width=0.9\linewidth]{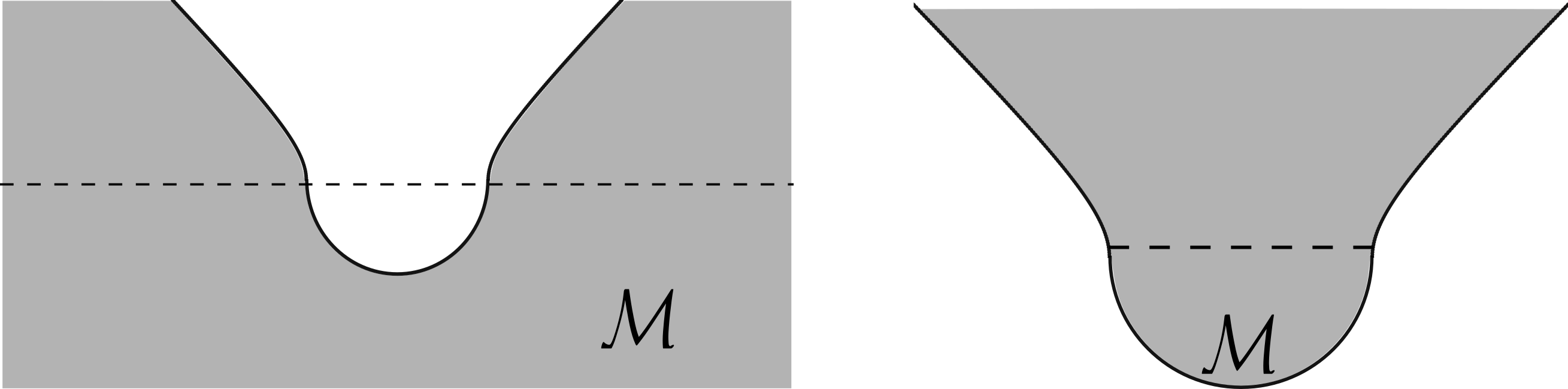}
	\caption{Physical processes corresponding to bubble of nothing (left) and bubble of something (right). The dashed line denotes the locus of analytic continuation from euclidean half instanton to the Lorentzian, expanding bubble.
    The solid black line is the boundary of 4d space.}%
	\label{fig_bubble_of_anything_Lorentzian}
\end{figure}
In the corresponding Lorentzian spacetime, a hole of nothing emerges as a quantum effect and then starts growing, eating up all space, as depicted on the l.h.s.~of Fig.~\ref{fig_bubble_of_anything_Lorentzian}.
This process and the corresponding instanton is known as a `bubble of nothing'.
Further studies of such decay processes considered models with stabilized extra dimensions as well as geometries where an originally present supersymmetry is broken by compactification on spheres \cite{Blanco-Pillado:2010xww,Blanco-Pillado:2016xvf,Brown:2010mf,Brown:2011gt,Blanco-Pillado:2010vdp,Dibitetto:2020csn,Draper:2021qtc,Draper:2021ujg,Draper:2023ulp} or in string-theoretic toy-models \cite{Brill:1991qe,Fabinger:2000jd,Horowitz:2007pr,Ooguri:2017njy,Acharya:2019mcu,GarciaEtxebarria:2020xsr,Bomans:2021ara}.

The reverse process of a decay to nothing is vacuum creation from nothing with an instanton of the type illustrated on the r.h.s.~of Fig.~\ref{fig_bubble_of_anything}.
The geometry is that of a disk with an $S^1$ over every point. At the boundary of the disk, the $S^1$ shrinks to zero size.
As the counterpart to the `bubble of nothing', we refer to the creation process as a `bubble of something'\footnote{In \cite{Blanco-Pillado:2011fcm}, the term `bubbles from nothing' is used.}. Again, these processes are different from the
most familiar vacuum creation proposals of \cite{Hartle:1983ai,Linde:1983mx,Vilenkin:1984wp}. Crucially, from the 4d perspective, the instanton contains a boundary and the cosmological constant of the new vacuum need not be positive.  Furthermore, the created universe is open in the sense that its spatial slices are negatively curved. The idea of a bubble of something was first introduced in \cite{Hawking:1998bn,Turok:1998he} and then connected to extra dimensions in \cite{Garriga:1998ri}. 
Further research on this topic includes \cite{Garriga:1998tm,Bousso:1998pk,Brown:2011gt,Blanco-Pillado:2011fcm,Cespedes:2023jdk}.

From the four-dimensional perspective, a bubble of nothing looks like an expanding end-of-the-world (ETW) brane with negative tension. The negativity of the tension is enforced by the Israel conditions, which relate the extrinsic curvature of the boundary to the tension of the ETW brane. Ordinary domain walls with negative tension typically suffer from instabilities, since it is energetically favorable to increase their volume. However, it is well known that ETW branes can consistently have negative tension. Examples are provided by O8 planes and the IR-region of the Klebanov-Strassler-throat \cite{Klebanov:2000hb}, which in turn can be viewed as a stringy realization \cite{Brummer:2005sh} of the IR-brane of the Randall-Sundrum model \cite{Randall:1999ee}. By contrast, a bubble of something requires a positive-tension ETW brane since, as can be seen by comparing the r.h.s. and l.h.s. of Fig. \ref{fig_bubble_of_anything}, the orientation of the boundary and hence the sign of the extrinsic curvature scalar is opposite.

\begin{figure}[ht]
	\centering
	\includegraphics[width=0.3\linewidth]{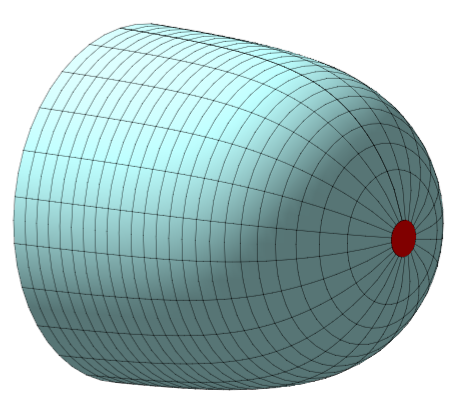}
	\caption{A bordism between $S^1$ and nothing, including a defect (drawn as a red disk).}%
	\label{fig_defect}
\end{figure}

The Witten bubble of nothing can not be naively generalized to supersymmetric configurations since only spinors with anti-periodic boundary conditions on the $S^1$ can be defined on the cigar geometry of Figure \ref{fig_bubble_of_anything}.
The reason is that a circle with periodic boundary conditions for the spinor is not in the trivial class of the spin cobordism group $\Omega_1^{Spin}=\mathbb{Z}_2$. As a consequence, the supersymmetric vacuum does not decay to nothing via the smooth bubble of nothing \cite{Witten:1981gj}.
More recently, however, it has been argued that this argument for stability is not quite correct in a fundamental theory of quantum gravity. Namely, according to the cobordism conjecture \cite{McNamara:2019rup} any internal manifold of a string compactification must be cobordant to nothing. In other words, there can not be a topological obstruction to letting spacetime end by the shrinking of the compact space to zero size. Consequently, defects ensuring the triviality of compact spaces in cobordisms are predicted \cite{McNamara:2019rup,Montero:2020icj,Dierigl:2022reg,Debray:2023yrs}.
A bordism with defect is illustrated in Fig.~\ref{fig_defect}. 
From the 4d point of view, the conjecture implies that all string compactifications possess ETW branes, but it remains a question of dynamics whether bubbles of nothing or something (bubbles of anything for short) arise. A key dynamical ingredient is the tension of the defect. In particular, one expects that in the case of supersymmetric compactifications the relevant defects are always heavy enough to ensure stability.
Other applications of the cobordism conjecture can be found in \cite{Buratti:2021fiv,Buratti:2021yia,Angius:2022aeq,Angius:2022mgh,Andriot:2022mri,Blumenhagen:2022bvh,Blumenhagen:2022mqw,Blumenhagen:2023abk}.

In what follows, we will develop a 4d effective field theory approach to vacuum decay/creation via bubbles of anything, and use it extensively. Let us illustrate this by discussing the Witten bubble, allowing for a defect as depicted in Fig.~\ref{fig_defect}. In this case (which of course is just a toy model), the defect has co-dimension 2 in the 5-dimensional spacetime,
and its tension is equivalent to a deficit angle $\theta$ of the shrinking circle.
The relevant 4d Lagrangian consists of the Einstein-Hilbert term and the kinetic term for a scalar $\phi$, characterising the radius of the $S^1$. In addition, there is a boundary action including the Gibbons-Hawking-York term and the tension of our ETW brane. Our analysis of this setting is related to\cite{Dine:2004uw,Draper:2021qtc,Draper:2021ujg}. It goes beyond these references by allowing for a defect. In addition, we differ in the treatment of the total derivative and boundary terms: We remove the former by integration, obtaining a standard 4d EFT with a completely local description of the ETW brane.

Boundary conditions at the brane fix the scalar field to a certain value, corresponding from the 5-dimensional perspective to a characteristic finite size, $\eta$, of the defect. The deficit angle $\theta$ is microscopically determined by the tension of the defect. In the 4d EFT, it is identified with the gradient with which $\phi$ approaches the ETW brane.
We will show that $\eta$ and $\theta$ are the key input needed for a simple 4d derivation of the instanton solution corresponding to the Witten bubble with defect and its action, which is related to the decay/creation rate in the usual way.

It can be shown that in the 5d model it is mathematically consistent to view the defect as being point-like, i.e.~to perform the limit $\eta\to 0$. In this case, $\phi$ runs to infinity at the ETW brane.
The original Witten bubble without a defect can now be obtained by setting $\theta = 0$ and performing the above limit $\eta\to 0$. However, even in this case it is useful to keep $\eta$ finite since it now plays the role of a regulator making a pure 4d Einstein-frame description possible.

As remarked above, defects of finite size and tension become intrinsic for string theory models such as Calabi-Yau orientifold compactifications with flux. Apart from that, the discussion of the two previous paragraphs carries over almost word by word. The CY is assumed to shrink at the ETW boundary. A defect of size $\eta$, which in contrast to the Witten case is now unavoidable, hides the appearing singularity. An appropriately generalized deficit angle $\theta$ characterizes the impact of the defect on the nearby geometry, and contributes to the tension of the ETW brane. While instanton solution and action can be given rather explicitly, the limit $\eta\to 0$ can not be taken. We expect that topology can fluctuate freely at string scale, so $\eta\sim 1$ in string units is plausible, but clearly also much larger ETW defects can exist.

Having thus developed an effective method for computing bubble-of-anything creation/decay rates, we turn to the 
construction of an ETW brane for type IIB CY orientifold compactifications with D3/D7 branes. This accounts for many vacua of the best understood part of the string landscape \cite{Denef:2004ze,Kachru:2003aw,Balasubramanian:2005zx,Conlon:2005ki}. 
The defect corresponding to our proposed ETW brane is an O5/D5 system wrapping a three-cycle of the CY. It can be understood as the mirror dual of a type-IIA orientifold model where, rather obviously, the world can end on an O8/D8 system wrapping the entire mirror manifold and filling three non-compact dimensions.
We provide leading order estimates for the transition rates for such vacua and compare the decay rates to traditional decay mechanisms \cite{Coleman:1980aw,Kachru:2002gs,Freivogel:2008wm,deAlwis:2013gka}.

We emphasize that in distinction to the previous discussion based on a rather generic Calabi-Yau cobordism, the orientifold construction does not require the compact space to become small. This illustrates that in controlled situations large defects of size $\eta=R_{\it KK}\gg 1$ are possible. On the other hand, being supersymmetric, the O5/D5 system will lead to a flat ETW brane and the associated transition rate will vanish at face value. 
However, including SUSY breaking and other corrections, it is also possible that our proposed ETW brane is driven to the previously discussed regime of small $\eta$.

The paper is organized as follows:
In Sec. \ref{sect_dynamics}, we set the stage by reviewing the Witten bubble and its 4d description.
In Sec. \ref{sect_generalized_Witten}, we present the Witten bubble of nothing with defect and characterize it in 4d language. Then we use the 4d intuition gained to analyze more general bubbles of anything, including CY compactifications of string theory.
In Sec. \ref{sect_landscape_applications}, we construct an ETW brane for vacua arising from compactifying type IIB string theory on a CY orientifold. 
We use Sec. \ref{sect_comments} to defend the bubble of something against fundamental criticism raised in \cite{Brown:2011gt} and to 
discuss the potential for making our analysis more precise.

\section{Bubbles of anything -- technical preliminaries}\label{sect_dynamics}
In this section, we first review the Witten bubble of nothing and then discuss the general setup for analyzing bubbles of anything from the four-dimensional perspective.

\subsection{The Witten bubble}
\label{twb}
Witten \cite{Witten:1981gj} has discovered that the Kaluza-Klein (KK) geometry $\mathbb{M}_4\times S^1$ exhibits a peculiar instability.
The relevant gravitational instanton is precisely the five-dimensional Schwarzschild black hole geometry, 
with metric
\begin{align}
    ds^2=\left(1-\frac{R_{KK}^2}{\tilde{r}^2}\right)^{-1}d\tilde{r}^2+\tilde{r}^2d\Omega_3^2+\left(1-\frac{R_{KK}^2}{\tilde{r}^2}\right)R_{KK}^2dy^2\,,\quad y\sim y+2\pi\,,\label{Witten_bon_5d_geometry}
\end{align}
satisfying Einstein's equations.
Here, we have chosen to name the radial coordinate $\tilde{r}$ instead of $r$ to reserve the latter for later use.
At $\tilde{r}\gg R_{KK}$, this geometry approaches the KK vacuum $\mathbb{M}_4\times S^1$ with circle radius $R_{KK}$.
Crucially, the geometry is defined only for $\tilde{r}\geq R_{KK}$ and at $\tilde{r}=R_{KK}$ the internal circle shrinks to zero size (cf.~the l.h.~side of Fig.~\ref{fig_bubble_of_anything}).
From the four-dimensional point of view, this locus represents the end of the world.
Despite the vanishing radius of the $S^1$, the five-dimensional curvature remains small as long as $R_{KK}$ is large in 5d Planck units, ensuring the reliability of a semiclassical analysis.

The Lorentzian geometry of the bubble of nothing is obtained by an analytic continuation of \eqref{Witten_bon_5d_geometry}: 
One first writes
\begin{align}
    d\Omega_3^2=d\psi^2+\sin^2\psi d\Omega_2^2\label{analytic_continuation}
\end{align}
and then lets $\psi\to \frac{\pi}{2}+it$, where $t\in(0,\infty)$. 
This step is analogous to the analytic continuation employed in Coleman-de Luccia instantons \cite{Coleman:1980aw}. From the 4d perspective, a spherical hole of nothing with radius $R_{KK}$ nucleates at $t=0$, starts accelerating and rapidly approaches the speed of light. This Lorentzian geometry corresponds to the region above the dashed line on the l.h. side of Fig.~\ref{fig_bubble_of_anything_Lorentzian}.

The nucleation probability per time and volume of such a bubble can be estimates as
\begin{align}
    \Gamma\sim \exp(-B)\,, \qquad B=S_{instanton}-S_{vacuum}\,.\label{decay_rate_general}
\end{align}
Here $S_{instanton}$ and $S_{vacuum}$ are defined by evaluating the Einstein-Hilbert action supplemented by the Gibbons-Hawking-York term,\footnote{We set the 5d Planck mass $M_5=1/8\pi G_5$ to unity. Moreover, our sign convention is such that $\mathcal{K}_5=\nabla_an^a$, with $n=n^a\partial_a$ being the outward pointing normal vector of the boundary hypersurface. We adopt this convention throughout the paper.}
\begin{align}
    S=-\frac{1}{2}\int\sqrt{g}\mathcal{R}_5-\int_{\partial \mathcal{M}}\sqrt{h}\mathcal{K}_5\,,
\end{align}
on the instanton geometry \eqref{Witten_bon_5d_geometry} and in the vacuum. In the above, $\mathcal{R}_5$ and $\mathcal{K}_5$ denote the 5d Ricci- and extrinsic curvature scalar respectively. One explicitly finds\cite{Witten:1981gj}
\begin{align}
    B=2\pi^3R_{KK}^3\,.\label{B_Witten}
\end{align}

\subsection{Bubbles of anything from the four-dimensional point of view}\label{sect_4d_view}
We want to find a 4d description of the Witten bubble and then study more general bubbles of nothing as well as bubbles of something from a 4d point of view. Our primary interest will eventually be in bubbles related to Calabi-Yau (CY) compactifications of string theory. 
We start by considering compactifications from $d=4+n$ to 4 dimensions on an $n$-dimensional compact manifold. To obtain the simplest possible estimate for rates of bubble-of-anything transitions, we restrict attention to the volume modulus. This is also motivated by the fact that in many string models the volume modulus has the lowest mass. We neglect all other fields in the following. As shown in Appendix \ref{appenndix_sect_dimensional_reduction}, the $d$-dimensional-Einstein-Hilbert action with metric ansatz
\begin{align}
    ds^2=e^{2\alpha\phi}ds_4^2+e^{2\beta\phi}ds_n^2\label{Einstein_frame_ansatz}
\end{align}
reduces to a 4d theory of gravity with Einstein-frame metric $ds_4^2$ and a canonically normalized volume modulus field $\phi$ if the constants $\alpha,\beta$ are chosen as
\begin{align}
    \alpha^2=\frac{n}{2(n+2)}\,,\qquad \beta=-2\frac{\alpha}{n}\,.\label{alpha_beta_equations}
\end{align}
Here $\phi$ is measured in 4d Planck units. In the following, we choose $\alpha<0$ and, correspondingly, $\beta>0$.
In Eq.~\eqref{Einstein_frame_ansatz}, $ds_n^2$ describes the geometry of the internal manifold with the convention that its volume is fixed to unity in $d$-dimensional Planck units. For later convenience, we define the quantity $R$ as
\begin{align}
    2\pi R=e^{\beta\phi}\,,\label{R_def}
\end{align}
which may be thought of as the radius of the internal manifold, also measured in $d$-dimensional Planck units.
Clearly, the case $n=1$ is appropriate for the Witten bubble. 

Motivated by general arguments \cite{Coleman:1977th} and similarly to \cite{Draper:2021qtc,Draper:2021ujg}, we make an $O(4)$-symmetric ansatz for the instanton geometry:
\begin{align}
    ds_4^2=dr^2+f(r)^2d\Omega_3^2\quad,\qquad \phi=\phi(r)\,.
\end{align}
For completeness, we also record the corresponding $d$-dimensional metric:
\begin{align}
    ds^2=e^{2\alpha\phi(r)}\left(dr^2+f(r)^2d\Omega_3^2\right)+e^{2\beta\phi(r)}ds_n^2\,.\label{d_dim_geometry}
\end{align}
The KK radius is
\begin{align}
    R_{KK}=\frac{e^{\beta\phi_0}}{2\pi}\,,\label{R_kk}
\end{align}
with $\phi_0$ the value of the scalar $\phi$ when the geometry approaches the vacuum.
The equations of motion, i.e. the 4d Einstein equations and the field equation for $\phi$ take the form
\begin{align}
    \phi''&+3\frac{f'}{f}\phi'=\frac{dV(\phi)}{d\phi}\,,\label{phi_eom}\\
    f'^2&=1+\frac{f^2}{3}\left(\frac{\phi'^2}{2}-V(\phi)\right)\,,\label{f_eom}
\end{align}
where primes denote derivatives with respect to $r$. A potential $V(\phi)$ can arise for example from internal curvature or fluxes.

A flat-space bubble of nothing instanton is a solution to the field equations 
where $\phi=$const. and $f'(r)\to 1$ at $r\to\infty$ and where $\phi\to -\infty$ at $r=0$. This implies that we chose $r$ to vanish at the bubble wall, which has radius $f(r=0)$.

For a bubble of something instanton \cite{Garriga:1998ri}, we may choose $r=0$ to be at the center of the bubble (see r.h.s.~of Fig.~\ref{fig_bubble_of_anything}), where 
$\phi$ takes some finite value and $f=0$. The boundary is then at some finite $r>0$, where $\phi\to -\infty$ and the value of $f$ characterizes the radius of the bubble.

Solutions to \eqref{phi_eom},\eqref{f_eom} involve three integration constants. 
Two of them are determined by boundary conditions for $\phi$ and $\phi'$ at the bubble wall.
The final parameter is then determined by requiring $\phi'$ to vanish at $r=0$ in the bubble of something case and at $r\to \infty$ in the bubble of nothing case.\footnote{
To 
be more precise, in the bubble of nothing case with vanishing potential, $V(\phi)\equiv 0$, all solutions share the feature that $\phi'$ vanishes at $r\to \infty$. In this case, the 3rd integration constant can be used to set the asymptotic value of $\phi$, fixing $R_{KK}=\exp(\beta\phi_0)$.
}

We now specialize our discussion to bubble of nothing solutions and focus on internal manifolds without scalar curvature or fluxes, e.g. fluxless CY or torus compactifications, such that the potential vanishes identically.
These are merely simplifying assumptions and we discuss their severity in Section \ref{sect_comments}.
A more detailed discussion of bubbles of something is given in Section \ref{sect_bubble_of_something}.
With the assumptions stated above, we can work with analytic solutions. 
Equation \eqref{phi_eom} can be integrated to give
\begin{align}
    f^3\phi'=C=const.\label{phi_prime_solution}
\end{align} 
Substituting the result \eqref{phi_prime_solution} into \eqref{f_eom} leads to
\begin{align}
    f'^2=1+\frac{C^2}{6f^4}\,,\label{f_prime_squared}
\end{align}
which can be used to obtain \cite{Draper:2021qtc}\footnote{Here we have chosen the positive root of \eqref{f_prime_squared} such that $f$ increases with growing $r$.}
\begin{align}
    \phi(f)=\phi_0-\frac{\sqrt{6}}{2}\text{arcsinh}\left(\frac{C}{\sqrt{6}f^2}\right)\,.\label{phi_f}
\end{align}
For $C>0$, we find that $\phi$ shrinks when approaching the bubble core.
Deriving the profiles $\phi(r)$ and $f(r)$ analytically is more complicated, see \cite{Draper:2021qtc,Draper:2021ujg}.
However, it is clear from \eqref{f_eom} that $f$ is a monotonically growing function of $r$, such that knowing the profile $\phi(f)$ should be sufficient for practical calculations.
In the limit of $f\to\infty$, $\phi$ approaches $\phi_0$ such that the definition of $\phi_0$ in \eqref{phi_f} is compatible with Equation \eqref{R_kk}.
Further, $f'$ approaches unity and spacetime becomes asymptotically Minkowski, as desired.
In the limit $f\to 0$, $\phi$ approaches $-\infty$ such that the internal $S^1$ shrinks to zero size.
As we will shortly see, $C$ can be fixed by looking at the boundary conditions on $\phi'$ at the bubble surface. The third integration constant is fixed implicitly by choosing the point where $\phi$ diverges to be at $r=0$.

From \eqref{phi_f} one immediately sees that $f=0$ at the bubble wall, where $\phi=-\infty$. This means that the bubble has vanishing radius in 4d Planck units. While this may sound confusing, it is simply a consequence of the fact that, at the wall, the compact space shrinks and hence the 4d Planck length diverges. This technical problem of the 4d Einstein frame analysis is overcome when, further down, we regulate the 4d solution by cutting it off when the compact space reaches minimal size $\eta$. If appropriate, $\eta$ may be identified with the size of an ETW defect.

\section{Generalized Witten bubbles}\label{sect_generalized_Witten}
We now turn to one of our central points: The development of an entirely four-dimensional description of bubble of nothing/something instantons related to the Witten bubble.
In Sec.~\ref{sect_modified_Witten}, we study the Witten bubble of nothing including a defect at the bubble wall.
In Sec.~\ref{sect_decay_rate}, we develop a method to effectively compute
decay rates for such bubbles, which will also apply to even more general bubble geometries to be discussed below. Section~\ref{sect_generalizations_and_exotic_DW} discusses a generalizion of the Witten bubble with defect to higher dimensions.
In Sec.~\ref{sect_CY}, bubbles of nothing in 10d Calabi-Yau (CY) compactifications in string theory are analyzed under the assumption that the instanton geometry is similar to the Witten bubble with defect.
Finally, in Sec. \ref{sect_bubble_of_something}, a more detailed discussion of bubbles of something is given.

\subsection{Witten bubble with defect}\label{sect_modified_Witten}
We use the analytic solution from the end of Sec.~\ref{sect_4d_view}, focussing on the case $d=5$ where
\eqref{alpha_beta_equations}
\begin{align}
    \alpha = -\sqrt{\frac{1}{6}}\,,\qquad \beta=\sqrt{\frac{2}{3}}\,.\label{alpha_beta_5d}
\end{align}
The integration constant $\phi_0$ is straightforwardly determined by demanding a specific KK radius at $f\to\infty$, cf. \eqref{R_kk}.
The integration constant $C$ remains to be fixed, which is done by analyzing the boundary conditions near the bubble wall.
For the geometry \eqref{d_dim_geometry} with $ds_n^2=dy^2\,, y\sim y+1$ to be completely regular, one has to demand that the $S^1$ shrink to zero size  without creating a deficit angle.
We can however be more general and consider the class of bubbles of nothing where an extra defect is needed to close off the geometry, cf.~Fig.~\ref{fig_defect}. In this case, a deficit angle generically arises. 
For related discussions of defects see \cite{Horowitz:2007pr,Blanco-Pillado:2010vdp,Blanco-Pillado:2010xww,Blanco-Pillado:2016xvf,Bomans:2021ara}. 

Since we work semiclassically, when we approach the defect we can only trust the geometry until the internal circle has shrunk to Planck size. If the defect is larger than Planck size, we have to stop even earlier. Let us denote the minimal value of $R$ up to which we trust our analysis by $\eta$, which for the moment we assume to be an $\order{1}$ number. In other words, the geometry stays trustworthy until a minimal $f=f_{\eta}$, defined by
\begin{align}
    \eta =R(f_{\eta})\,.\label{eta_def_5d}
\end{align}
Assuming $R_{KK}\gg \eta$ and using \eqref{R_def},\eqref{phi_f}, we find
\begin{align}
    f_{\eta}=\left(\frac{2C}{\sqrt{6}}\right)^{1/2}\left(\frac{\eta}{R_{KK}}\right)^{1/2}\,.\label{f_eta_5d}
\end{align}

As already noted above our defect, being a 3d object in 5d, creates a deficit angle in the geometry. This is entirely analogous to a point mass in three or a cosmic string in four dimensions. We can thus relate the defect tension $T_{def}$ to the deficit angle and hence to the local geometry in the standard way \cite{Garfinkle:1985hr,Vilenkin:1984ib},
\begin{align}
    T_{def}=\theta \,\,\,,\qquad
    1-\frac{\theta}{2\pi}=\left.\frac{dR}{dx}\right|_{x=0}\,.\label{deficit_angle_def}
    \end{align}
Here we use 5d Planck units and $x$ measures the proper distance from the defect. It is related to $r$ by
\begin{align}
    dx=e^{\alpha\phi(r)}dr\,,\label{dx_dr_rel}
\end{align}
such that one finds
\begin{align}
    1-\frac{\theta}{2\pi}=\left.\frac{\beta}{2\pi}\phi'e^{(\beta-\alpha)\phi}\right|_{x=0}=\left.\frac{\beta}{2\pi}\frac{C}{f^3}e^{(\beta-\alpha)\phi}\right|_{f=f_{\eta}}=\sqrt{\frac{4\pi}{3}}\frac{C}{f_{\eta}^3}\eta^{3/2}\,.\label{theta_def_5d}
\end{align}
Here, we have used the solution \eqref{phi_prime_solution}, the definition of $\eta$ \eqref{eta_def_5d} and \eqref{alpha_beta_5d}.

After substituting \eqref{f_eta_5d} into \eqref{theta_def_5d}, one arrives at
\begin{align}
    C=\sqrt{6}\pi R_{KK}^3\left(1-\frac{\theta}{2\pi}\right)^{-2}\,.\label{C_5d}
\end{align}
Interestingly, $C$ does not depend on $\eta$ which reflects the fact that, even though a defect is present, the limit $\eta\to 0$ may be taken. In other words, it is mathematically (though maybe not physically) consistent to take the defect to be point-like in the two transverse directions. We will see below that this is no longer true in higher-dimensional models.

We have now fully determined the solutions corresponding to the Witten bubble with defect from the four-dimensional perspective by demanding the three boundary conditions \eqref{R_kk},\eqref{eta_def_5d},\eqref{theta_def_5d}.
By inserting \eqref{C_5d} into \eqref{phi_f}, one obtains the instanton solution depending only on the input parameters $R_{KK}$ and $\theta$ and where the limit $\eta\to 0$ has implicitly been taken.
The Witten bubble is recovered for the special case $\theta=0$. The critical bubble radius is
\begin{align}
    \rho_0=\left.e^{\alpha\phi}f\right|_{f=f_\eta}=R_{KK}^{-1/2}\left(\frac{1}{\pi\sqrt{6}}\right)^{1/2}\sqrt{C}=\frac{R_{KK}}{1-\frac{\theta}{2\pi}}\,.\label{rho0_5d}
\end{align}
We observe that $\rho_0$ matches the Witten result for $\theta=0$ and becomes larger when turning on a deficit angle. 
This agrees with our physics intuition that we expect an additional defect to result in a larger critical bubble and a smaller decay rate.

\subsection{The decay rate}\label{sect_decay_rate}
We are now ready to calculate decay rates for bubbles of nothing with defects. Although we have the Witten bubble with defect in mind as our primary example, our analysis will turn out to be valid for more general bubbles of nothing, such as the 10d string theory models we are ultimately after.
We follow the strategy explained at the end of Sec.~\ref{twb}, computing in particular the decay exponent using \eqref{decay_rate_general}. Crucially, we want to work with 4d effective actions for the instanton and the vacuum contribution. Both take the form
\begin{align}
    S=\int_{\mathcal{M}} \sqrt{g}\left(-\frac{1}{2}\mathcal{R}_4+\frac{1}{2}(\partial \phi )^2+V(\phi)\right)-\int_{\partial \mathcal{M}} \sqrt{h}\left(\mathcal{K}_4-T_4\right)\,.\label{S4_GHY}
\end{align}
Here, $\mathcal{R}_4$ is the 4d Ricci scalar, $\mathcal{K}_4$ is the extrinsic curvature, $\partial \mathcal{M}$ is the boundary of the four-dimensional spacetime $\mathcal{M}$ with induced metric $h$. For both $S_{instanton}$ and $S_{vacuum}$, there is a boundary contribution at infinity, without brane tension $T_4$. These contributions will cancel. At the center, the vacuum obviously has no boundary while $S_{instanton}$ receives a contribution with both non-zero $\mathcal{K}_4$ and the tension $T_4$ of an effective 4d ETW brane sitting at the bubble wall, i.e.~at the spherical boundary of $\mathcal{M}$ on the l.h.s. of Fig.~\ref{fig_bubble_of_anything}.
The subscript $4$ in $T_4$ indicates the dimension of space which is bounded by the ETW brane and should not be confused with the tension of a $4$-brane in string theory.
In Eq.~\eqref{S4_GHY}, the 4d Planck mass is set to unity. It is hence important to distinguish between four- and $d$-dimensional quantities, which are measured in the respective Planck units.

The form of the action \eqref{S4_GHY} differs from the one used in \cite{Dine:2004uw,Draper:2021qtc,Draper:2021ujg} by the absence of a term proportional to $\square \phi$ coming from dimensional reduction of $\mathcal{R}_d$.  We have replaced this total derivative term by appropriate boundary terms: One at infinity and one at the inner spherical boundary. At the former boundary, the term coming from $\square \phi$ is cancelled by a contribution deriving from the 5d GHY term.
The latter boundary is the ETW boundary defined by $f=f_{\eta}$, where also the ETW brane with tension $T_4$ sits. Here, the boundary term coming from $\square \phi$ is absorbed in our effective description in terms of $\mathcal{K}_4$ and $T_4$.
We feel that such a description, based on \eqref{S4_GHY} without total derivative terms but viewing the bubble surface as an additional boundary, is more appropriate for our subsequent stringy applications with physical defects at the bubble wall.

In order to compute the on-shell action of the instanton, we first note that the trace of the 4d Einstein equations determines
\begin{align}
    \frac{1}{2}\mathcal{R}_4=\frac{1}{2}\phi'^2+2V(\phi)\,,
\end{align}
and hence
\begin{align}
    S_{\text{on-shell}}=-2\pi^2\int dr\,f^3V(\phi)-\int_{\partial \mathcal{M}}\sqrt{h}\left(\mathcal{K}_4-T_4\right)\,.\label{S_int_on_shell}
\end{align}
In the absence of a potential, the bulk contribution simply vanishes and the entire contribution comes from boundary terms. 

For a Minkowski bubble of nothing geometry, there are two boundaries: One located at asymptotic infinity $f\to\infty$ and one near the center of the bubble at finite $f=f_{\eta}$. 
We calculate the value of the extrinsic curvature scalar for the boundary at infinity using \eqref{f_eom}
\begin{align}
    \mathcal{K}_4=3\frac{f'}{f}=\frac{3}{f}\sqrt{1+\frac{C^2}{6f^4}}\simeq \frac{3}{f}+\frac{C^2}{4f^5}\,,
\end{align}
where in the last step, we have expanded the previous expression to subleading order in the limit $f\to\infty$.
The contribution from this boundary to the decay exponent $B$ (see \eqref{decay_rate_general}), can be computed by noting that the vacuum extrinsic curvature scalar is given by 
\begin{align}
    \mathcal{K}_{4,vacuum}=\frac{3}{f}\,,
\end{align}
such that the contribution to $B$, i.e. the difference in the actions, can be calculated
\begin{align}
    B\supset \int_{f\to\infty}f^3\left(-\mathcal{K}_4+\mathcal{K}_{4,vacuum}\right)=6\pi^2f^3\left.\left(-\frac{3}{f}-\frac{C^2}{4f^5}+\frac{3}{f}\right)\right|_{f\to\infty}=0\,.
\end{align}
Hence, the entire contribution to $B$ comes from the ETW boundary at the bubble wall.
As dictated by the Israel junction conditions, an ETW boundary with non-vanishing extrinsic curvature requires a brane with a certain tension.
The (analogue of the) Israel conditions relate the extrinsic curvature $\mathcal{K}_{4,ab}$ to the matter energy momentum tensor $S_{ab}$ located at the end of the world \cite{Takayanagi:2011zk,Fujita:2011fp}
\begin{align}
    \mathcal{K}_{4,ab}=S_{ab}-\frac{1}{2}h_{ab}h^{cd}S_{cd}\,.\label{Israel_condition}
\end{align}
An object of pure tension $T_4$ has an energy momentum tensor $S_{ab}=-T_4h_{ab}$ and, due to the $O(4)$ symmetry of the problem, we lose no information by taking the trace of \eqref{Israel_condition}
\begin{align}
    \mathcal{K}_4=-\frac{1}{2}h^{ab}S_{ab}=\frac{3}{2}T_4\,.\label{K_T_relation}
\end{align}
As a consequence of \eqref{K_T_relation}, the decay exponent takes the form\footnote{For the bubble of nothing solution, where the action of the vacuum has to be subtracted, we note that the vacuum action does not have a boundary at finite $f=0$. However, if one considers $f=0$ to be a boundary of flat space one quickly notices that the extrinsic curvature reads $\mathcal{K}_{4,vacuum}=-3/f$ and the boundary contribution becomes $6\pi^2f^3/f|_{f=0}=0$.}
\begin{align}
    B= 2\pi^2 f_\eta^3\left(-\mathcal{K}_4+T_4\right)=-2\pi^2f_\eta^3\frac{T_4}{2}=-2\pi^2f_\eta^3\frac{\mathcal{K}_4}{3}\,.\label{B_bdry_contr}
\end{align}
Finally, for the ETW boundary at $f_{\eta}$, the extrinsic curvature scalar takes the value
\begin{align}
    \mathcal{K}_4=- 3\left.\frac{f'}{f}\right|_{f=f_{\eta}}=-\frac{3}{f_{\eta}}\sqrt{1+\frac{C^2}{6f_{\eta}^4}}\,.\label{K4_wall}
\end{align}
By comparing with \eqref{K_T_relation}, we see that a Minkowski bubble of nothing requires a negative tension $T_4$, as anticipated before.

We can now compute the decay rate for the Witten bubble with defect.
Using \eqref{K_T_relation}, \eqref{K4_wall}, \eqref{C_5d} in the limit $\eta\ll R_{KK}$, the tension and extrinsic curvature are calculated as
\begin{align}
    T_4=\frac{2}{3}\mathcal{K}_4=-\frac{2C}{\sqrt{6}f_{\eta}^3}=-\left(1-\frac{\theta}{2\pi}\right)\frac{\eta^{-\frac{3}{2}}}{\sqrt{2\pi}}\,.\label{K4_5d}
\end{align}
The decay exponent then follows from substituting \eqref{f_eta_5d},\eqref{K4_5d} into \eqref{B_bdry_contr}
\begin{align}
    B=\frac{2\pi^3R_{KK}^3}{\left(1-\frac{\theta}{2\pi}\right)^2}=\frac{\pi^2M_P^2R_{KK}^2}{\left(1-\frac{\theta}{2\pi}\right)^2}\,,\label{B_Witten_modified}
\end{align}
where we have used the definition of the 4d Planck mass $M_P^2=2\pi R_{KK}$.
We observe that the Witten bubble decay rate \eqref{B_Witten} is recovered for $\theta=0$.
Furthermore, the decay rate decreases for increasing deficit angle (and hence increasing defect tension) and the transition becomes forbidden for $\theta=2\pi$. 
By comparing with \eqref{rho0_5d}, we see that in this limit, the bubble radius grows infinitely large.

Let us collect some further remarks on the above calculation. 
From \eqref{K4_5d}, we see that the Witten bubble solution has, from the four-dimensional perspective, a non-vanishing extrinsic curvature contribution to the action at the bubble surface, which for a low-energy observer signals the presence of a boundary ETW brane.
Hence, the description using $\mathcal{K}_4$ and $T_4$ is justified.
In the presence of more complicated defects as discussed below, it is likely that even the $d$-dimensional theory is described by a boundary action at the locus of the defect, which then naturally descends to a 4d boundary action as in \eqref{S4_GHY}.
Importantly, the action of the form \eqref{S4_GHY} also allows us to reliably compute decay rates for more general ETW branes, as will be discussed below.

From \eqref{K4_5d}, we see that $T_4$ diverges in the limit $\eta\to 0$.
This is result from $T_4$ being measured in units of the 4d Planck mass, which in the limit of $\eta\to 0$ locally vanishes as $\sim\eta^{1/2}$. 
Hence, the diverging $T_4$ is a peculiar looking artefact of our 4d description which however will turn out to be useful in the upcoming sections.

The additional defect was introduced in \eqref{deficit_angle_def} as an object of the 5d theory wrapping the $S^3$ at fixed $f$.
Its tension in 4d Einstein frame can be obtained by looking at the action
\begin{align}
    S_{def}=-T_{def}\int d^3x \sqrt{g_5|_{P.B.}}=-T_{def}\int d^3x e^{3\alpha\phi}f^3=-T_{def}e^{-\frac{3}{2}\beta\phi}\int d^3x \sqrt{g_4|_{P.B.}}\,,
\end{align}
where we have used \eqref{appendix_alpha_beta_rel}.
Hence, the tension in 4d Einstein frame is given by $T_{def}(2\pi R)^{-3/2}$.
Remembering that $T_{def}=\theta$, Eq. \eqref{K4_5d} becomes intuitive: The defect tension simply adds to the negative tension arising from the shrinking extra dimension and brings the total tension closer to zero, making the decay less likely to occur.

The positive energy theorem \cite{Witten:1981mf,Dai_2004,Dai_2005} applies to KK compactifications which asymptote to flat Minkowski space times constant KK manifold. If there exists an asymptotically covariantly constant spinor and the weak energy condition holds\footnote{There are multiple versions of the positive energy theorem with slightly different assumptions. For our purposes it is appropriate to use \cite{Dai_2004}.}, the ADM mass is non-negative. Furthermore, the mass only vanishes if the asymptotic vacuum situation extends to the complete geometry. Since energy is conserved, this implies that the vacuum is stable. Based on arguments of this type and using the fact that the energy is the square of the supercharges, one expects supersymmetric string vacua to be stable \cite{Giri:2021eob}. In our context, we thus expect $T_{def}$ to be large enough to forbid a decay if the relevant vacuum is supersymmetric. Once supersymmetry is broken and the weak energy condition is violated, decays become possible\cite{GarciaEtxebarria:2020xsr}.

\subsection{Generality of the method developed}\label{sect_generalizations_and_exotic_DW}
Our method for computing the decay rate allows us to go beyond the Witten bubble with defect and to treat more complicated instanton geometries, e.g.~with more compact dimensions. Using the general metric ansatz \eqref{d_dim_geometry} with appropriate values for $\alpha$ and $\beta$ \eqref{alpha_beta_equations}, the equations of motion \eqref{phi_eom},\eqref{f_eom} continue to hold. If the potential $V(\phi)$ vanishes, also the solution \eqref{phi_prime_solution},\eqref{phi_f} remains valid.
Of course, our ability to let the compact space shrink to zero size in such a way that the higher-dimensional geometry remains smooth is special to $S^1$ and higher spheres. If the compactification space is generic, the locus where its volume vanishes will represent a singularity. 
Motivated by the Cobordism Conjecture \cite{McNamara:2019rup}, we may expect that corresponding transitions to nothing are allowed in quantum gravity.
Put differently, one may expect that spacetime topology can fluctuate at small scales and that hence our compact space, once it has become small, can disappear. We should then be able to cut out a region with sufficiently small compact space and to replace it by an effective `ETW defect'. The relevant region, i.e. the thickness of the ETW defect, may be Planck size or it may be larger. This is to a certain extent a matter of definition of our effective theory.
Of course, for our approximation to be valid, the thickness should be much smaller than the critical bubble radius.

This class of instantons, where only the volume of the internal manifold changes until an ETW defect is reached, is conceptually similar to the Witten bubble with defect and might hence be viewed as a generalized Witten bubble.
Many known bubble of nothing solutions fall in this category and may hence be described by our general methods, see e.g. \cite{GarciaEtxebarria:2020xsr, Brown:2011gt,Draper:2021qtc}.
The power of our 4d description and the generality of the ansatz \eqref{S4_GHY} allow us to compute the decay rate for a general such instanton, i.e.~equations \eqref{B_bdry_contr},\eqref{K4_wall} remain valid. 

A natural limit to consider is $\eta\ll R_{KK}$. 
In this case one finds from  \eqref{B_bdry_contr},\eqref{K4_wall} that
\begin{align}
    B=\frac{2\pi^2}{\sqrt{6}}|C|\,.\label{decay_rate_BON_general}
\end{align}
Like for the Witten bubble, the integration constant $C$ depends on $R_{KK}$ and on the boundary conditions at the ETW defect. We will work this out below in examples.

Another natural class of solutions has $\eta\approx R_{KK}$. In this case, $1\gg C^2/f_\eta^4$, such that $f'\simeq 1$, $\phi'\simeq 0$ on the entire instanton. As a result, the ETW brane cuts out a sphere from flat Minkowski space. Moreover, the ETW defect is maximally thick and essentially identical with the ETW brane. We will see an explicit example in what follows.
The general formulas without any restriction on $\eta$ are shown in App.~\ref{appendix_sect_general_dimension}.

We note that, in principle, one could even have exotic bubble-of-nothing or something solutions in which the compact radius grows as one approaches the ETW brane. One may then have $\eta \gg R_{KK}$. Although such boundary conditions are physically unintuitive, the method how to compute the decay rate remains unchanged.

Another interesting observation to be made is that the extrinsic curvature \eqref{K4_wall} and hence the decay rate is symmetric under $C\to -C$.
From \eqref{phi_f}, one observes that this change in sign changes the solution from a shrinking to a growing internal manifold when approaching the bubble surface. 
The Witten bubble with defect is a solution with shrinking $S^1$, i.e. $C>0$, but due to $T$-duality, it might not be entirely surprising that an analogous solution with $C<0$ also exist.
Of course, the resulting geometry is somewhat peculiar, because the extra dimension grows large towards the bubble center and it is unclear how the geometry closes.
Under T-duality, the codimension-two defect assumed to be present gets mapped to a codimension-one brane wrapping the internal $S^1$. 
This brane can act as an ETW brane, i.e. fulfilling the Israel junction conditions, such that the geometry can smoothly end.
However, for theories in $d>5$, the T-duality interpretation is not always given. 
It would be interesting to study the $C\to -C$ symmetry in more detail.
\subsection{Bubbles of nothing in Calabi-Yau compactifications}\label{sect_CY}
In this section, we finally discuss bubble of nothing instanton solutions and the corresponding decay rates in string theory models arising from compactifiction on CY-orientifolds, possibly with fluxes.
We will use the intuition from the Witten bubble with defect and the general procedure how to calculate decay rates from above.
Fluxes and non-perturbative effects generally present in $\mathcal{N}=1$ or $\mathcal{N}=0$ models induce a potential for $\phi$ which is particularly important at small volumes. We assume that it is a valid approximation to treat the effects of this potential as being part of the effective ETW defect and hence use \eqref{phi_f} to find the decay rate.
This is justified as long as the thickness of the domain wall remains small compared to the critical bubble radius.

Since we now work in $d=10$, $\alpha$ and $\beta$ take values according to \eqref{alpha_beta_equations}:
\begin{align}
    \alpha=-\sqrt{\frac{3}{8}}\,,\qquad \beta=\frac{1}{2\sqrt{6}}\,.
\end{align}
Clearly, in the present case, we always require a `cobordism transition', i.e. a region of topology change and hence a non-trivial ETW defect.
Thus, we can trust the geometry defined by \eqref{phi_f} only up to the point where the CY radius has changed to some critical size $\eta$. 
From the condition
\begin{align}
    \eta=R(f_\eta)\,,\label{eta_def}
\end{align}
one then finds the relation 
\begin{align}
    \frac{C}{f_\eta^2}=\frac{\sqrt{6}}{2}\left[\left(\frac{R_{KK}}{\eta}\right)^4-\left(\frac{\eta}{R_{KK}}\right)^4\right]\,.\label{C_f2_CY}
\end{align}
As for the Witten bubble with defect, we expect that $C$ should be fixed by the boundary conditions at the ETW defect. While this ETW defect is an unfamiliar object, it is natural to characterize its effect on the surrounding geometry by a `deficit angle' $\theta$ defined in analogy to~\eqref{deficit_angle_def}:
\begin{align}
    1-\frac{\theta}{2\pi}=\left.\frac{dR}{dx}\right|_{x=0}=\left.\frac{\beta}{2\pi}\frac{C}{f^3}e^{(\beta-\alpha)\phi}\right|_{f=f_\eta}=\frac{1}{4\pi\sqrt{6}}\frac{C}{f_\eta^3}(2\pi\eta)^{4}\,.\label{theta_def_CY}
\end{align}

For the remainder of this section, we assume $\eta \ll R_{KK}$ and hence $C>0,\;\theta<2\pi$, which is the parameter range corresponding to a small ETW defect, as motivated above. Hence, we ignore the second term in \eqref{C_f2_CY}.
General equations, including the cases $\eta\approx R_{KK}$ and $\eta>R_{KK}$, are provided in App.~\ref{appendix_sect_general_dimension}.
Solving \eqref{C_f2_CY} and \eqref{theta_def_CY} for $C$ gives
\begin{align}
    C=\frac{\sqrt{6}(2\pi)^6}{32\left(1-\frac{\theta}{2\pi}\right)^2}\frac{R_{KK}^{12}}{\eta^4}\,.
\end{align}
This fully determines the instanton solution. Unlike the 5d case, $C$ now explicitly depends on $\eta$ and diverges as $\eta\to 0$. Thus, the limit of a truly point-like ETW defect can not be taken. To find the decay rate, we go on to compute the 4d extrinsic curvature of the boundary at $f=f_\eta$ (see \eqref{K4_wall}):
\begin{align}
    \mathcal{K}_4=-6\left|1-\frac{\theta}{2\pi}\right|(2\pi)^{-3}\eta^{-4}\,.
\end{align}
Note that $\mathcal{K}_4$ does not depend on $R_{KK}$, which in Appendix \ref{appendix_sect_general_dimension} is shown to hold generally in the limit $\eta\ll R_{KK}$. This is consistent with the expectation that the choice of $\eta$ and $\theta$ should completely fix the local geometry near the boundary.
By the Israel conditions, the extrinsic curvature also determines the total tension of the ETW brane in the 4d EFT, cf.~\eqref{K_T_relation}.

Next, we compute the critical bubble radius:
\begin{align}
    \rho_0=e^{\alpha\phi(f_\eta)}f_{\eta}=\frac{R_{KK}}{4\left|1-\frac{\theta}{2\pi}\right|}\left(\frac{R_{KK}}{\eta}\right)^3\,.\label{rho0_CY}
\end{align}
It diverges in the limit $\eta\to 0$, showing again that it is not possible to remove the ETW defect. In comparison to the Witten bubble radius \eqref{rho0_5d}, $\rho_0$ is much larger due to the appearance of the factor $(R_{KK}/\eta)^3\gg 1$.
The decay exponent straightforwardly follows from \eqref{decay_rate_BON_general}
\begin{align}
    B=\frac{(2\pi)^8R_{KK}^8}{64\left(1-\frac{\theta}{2\pi}\right)^2}\left(\frac{R_{KK}}{\eta}\right)^4=\frac{\pi^2M_P^2R_{KK}^2}{16\left(1-\frac{\theta}{2\pi}\right)^2}\left(\frac{R_{KK}}{\eta}\right)^4\,.\label{B_CY}
\end{align}
It is again enhanced relative to the Witten bubble case \eqref{B_Witten_modified}
by powers of $R_{KK}/\eta$.

A peculiar feature of the CY bubble is the fact that when approaching the bubble wall from infinity, the transversal three sphere radius $\rho=e^{\alpha\phi}f$ first shrinks as expected, but at some point starts growing again. 
This can be seen by looking at the profile $\rho(f)$
\begin{align}
    \rho(f)=R_{KK}^{-3}\left(\frac{C}{\sqrt{6}f^2}+\sqrt{\left(\frac{C}{\sqrt{6}f^2}\right)^2+1}\right)^{\frac{3}{4}}f\,.
\end{align}
For large values of $f$, $\rho(f)\approx R_{KK}^{-3}f$ and hence $\rho$ grows with growing $f$.
For small $f$ on the other hand,
\begin{align}
    \rho(f)\approx R_{KK}^{-3}\left(\frac{2C}{\sqrt{6}}\right)^{\frac{3}{4}}\frac{1}{\sqrt{f}}\,,
\end{align}
and we see that $\rho$ decreases with growing $f$. 
Such a behavior has already been observed in 6d models \cite{Blanco-Pillado:2010vdp}.
In Appendix \ref{appendix_sect_general_dimension}, it is shown that this peculiar behavior always occurs for $d>5$.

Clearly, this analysis could have been carried out with $C<0$ and from \eqref{decay_rate_BON_general} one can see that the decay rate is symmetric under $C\to-C$. In contrast to the case of $S^1$ compactifications, it is not immediately clear how to relate these two types of solutions using T-duality or mirror symmetry. It would be interesting to study this further, presumably invoking Calabi-Yau moduli other than the volume.

\subsection{Bubbles of something}\label{sect_bubble_of_something}
The 4d geometry corresponding to the bubble of something instanton is the disc on the r.h.s. of Fig.~\ref{fig_bubble_of_anything}, with $f=0$ at the center and $f=f_\eta$ at the boundary. To find the instanton solutions, we again need to consider the field equations \eqref{phi_eom},\eqref{f_eom} subject to the general boundary condition $\phi'(f=0)=0$ as well as the boundary conditions on $\phi$ and $\phi'$ at the bubble wall.
For instanton solutions in concrete 6d models, see \cite{Blanco-Pillado:2011fcm}.

Given a solution, one can calculate $-S_{instanton}$.
We would now like to compute a creation rate $\Gamma$, analogous to the decay rate of \eqref{decay_rate_general}. However, there is no parent vacuum and hence the natural definition of a rate is
\begin{align}
    \Gamma \sim \exp(-B)=\exp(-S_{instanton})\,,
\end{align}
without the subtraction of any parent vacuum action in the exponent.
This rate for the creation of a compact universe with boundary is completely analogous to the rate for the creation of a boundary-free, spherical universe suggested by Hartle and Hawking \cite{Hartle:1983ai}. The competing proposal by Linde and Vilenkin \cite{Linde:1983mx,Vilenkin:1984wp} famously differs by a sign in the exponent, but is similar in the aspect that no background action is subtracted.

Furthermore, the meaning of $\Gamma$ is altered in comparison to decay rates relating different vacua as discussed e.g.~in~\cite{Garriga:2005av,DeSimone:2008bq,Garriga:2012bc} (see also \cite{Friedrich:2022tqk} for a recent proposal): The rate no longer counts the number of creation events per time and volume since there is no background in which the bubble of something nucleates. Rather, only ratios of the form \be
\Gamma_i/\Gamma_j\,,
\ee
with $i,j$ labelling vacua of the theory, may be viewed as physical predictions. They specify the relative probabilities of creating vacuum $i$ rather then $j$ from nothing.  Such ratios would also be independent of the background action in the case of decay rates.

We first observe that, in general, bubbles of something do not exist for vanishing potential. This follows from \eqref{phi_prime_solution}, which shows that the vanishing of $f$ at the center, together with smoothness of $\phi$, implies $C=0$. This special solution corresponds to a flat disc with $\phi=\phi_0$. 

However, it is not difficult to show that bubble of something solutions do exist for nontrivial potentials.
They can approximate the flat-disc/constant-$\phi$ solution above as follows: In the central part of the disc, $\phi$ is essentially constant, sitting near a Minkowski-minimum of the potential. Near the boundary, $\phi$ starts moving away from that minimum and $f$ starts deviating from $f=r$, such that eventually the boundary conditions associated with the ETW defect are met at $f=f_\eta$.
The boundary conditions are given by \eqref{eta_def} and an equation for the deficit angle similar to \eqref{theta_def_CY}.
However, we have to remember that $x$ measures the distance from the ETW brane which, in comparison to \eqref{dx_dr_rel}, is now related to $r$ via
\begin{align}
    dx=-e^{\alpha\phi}dr\,.
\end{align}
As a result, the second boundary condition becomes
\begin{align}
    1-\frac{\theta}{2\pi}=\left.\frac{dR}{dx}\right|_{x=0}=-\left.\frac{\beta}{2\pi}\phi'e^{(\beta-\alpha)\phi}\right|_{f=f_\eta}=-\frac{1}{4\pi\sqrt{6}}\phi'(2\pi\eta)^{4}\,.\label{theta_def_CY_BoS}
\end{align}
Hence, a solution with deficit angle $\theta<2\pi$ now corresponds to $\phi'<0$ at the ETW brane, unlike in bubble of nothing models where $\phi'>0$ (i.e. $C>0$) was a necessary condition, cf. \eqref{theta_def_CY}.

The instanton action can be evaluated using our general formula \eqref{S_int_on_shell} for on-shell actions. It receives both bulk and boundary contributions. The boundary term contribution to $S_{instanton}$ can be computed using
\begin{align}
    \frac{3}{2}T_4 = \mathcal{K}_4=+\left.\frac{3f'}{f}\right|_{f=f_\eta}=+\frac{3}{f_\eta}\sqrt{1+\frac{f_{\eta}^2}{3}\left(\frac{\phi'^2}{2}-V(\phi)\right)_{f=f_\eta}}\geq 0\,.
\end{align}
Note that, as anticipated, a bubble of something requires a positive-tension ETW brane.

Given a solution, also the bulk term in \eqref{S_int_on_shell} can in principle be evaluated straightforwardly. However, in this study we do not discuss any details of the potential and we can hence only give explicit results if this term is negligible due to appropriate assumptions:

We can provide a creation rate estimate assuming that $\phi$ is stabilized such that $V \simeq 0$ for a large part of the instanton. Consequently, $\phi$ moves significantly only close to the ETW boundary.
Using the equations of motion, one can show that the bulk of the instanton is essentially flat space, i.e.~$\phi'\simeq 0, f'\simeq 1$.
In case of the critical bubble radius being large in comparison to the region where $\phi$ moves significantly, we may absorb these effects into the ETW brane tension and hence obtain an effective description of a bubble of something instanton where an ETW brane with tension $T_4$ bounds flat space, similar to \cite{Bousso:1998pk}.
The assumptions stated above are most easily fulfilled for small values of $T_4$ or a potential stabilizing $\phi$ strongly. Of course, by absorbing the unknown effects of the potential near the boundary into $T_4$, we sacrifice the calculability of $T_4$ from $\eta$ and $\theta$.

The extrinsic curvature at $f=f_\eta$, defined as the region where $\phi$ begins rolling and hence where the ETW brane is located, is given by
\begin{align}
    \mathcal{K}_4\simeq\frac{3}{f_\eta}\,.\label{K4_BoS_naive}
\end{align}
The ETW brane tension, now also capturing the effects of the potential and the rolling of $\phi$ near the ETW boundary, is then related to $f_\eta$ by \eqref{K_T_relation}
\begin{align}
    T_4=\frac{2}{3}\mathcal{K}_4=\frac{2}{f_\eta}.
\end{align}
Disregarding, as discussed, the bulk contribution and reinstating $M_P$ for clarity one finds
\begin{align}
    B=-2\pi^2 f_\eta^3 \frac{M_P^2\mathcal{K}_4}{3}=-2\pi^2f_\eta^2=-8\pi^2\frac{M_P^6}{T_4^2}\,.\label{B_BoS}
\end{align}

The rates governed by this exponent have to be viewed as competing with the famous proposals for vacuum creation \cite{Hartle:1983ai,Linde:1983mx,Vilenkin:1984wp} which result in closed universes and require the initial value of the scalar potential to be positive.
By contrast, the bubbles of something governed by \eqref{B_BoS} give rise to open universes with no restriction on the vacuum energy, as long as an ETW brane with sufficient tension exists.

Our Coleman-De Luccia based perspective  on \eqref{B_BoS} implies that the quantity in the exponent is eventually positive, such that no tunneling-suppression arises. This is then analogous to the Hartle-Hawking \cite{Hartle:1983ai} rather than the Linde/Vilenkin \cite{Linde:1983mx,Vilenkin:1984wp} prediction. However, in light of the long-lasting debate concerning these proposals, one should obviously ask whether a Linde-Vilenkin-like exponential suppression by $B$ from \eqref{B_BoS} may also be argued for.

Recently, the creation of universes from nothing according to \cite{Hartle:1983ai,Linde:1983mx,Vilenkin:1984wp} was analyzed based on the Lorentzian path integral \cite{Feldbrugge:2017fcc,Feldbrugge:2017mbc,Feldbrugge:2018gin,Vilenkin:2018dch,Vilenkin:2018oja}, where it was found that neither proposal in the original form is free of issues. An important role is played by potentially unsuppressed field fluctuations. Clearly, it would be desirable to study the bubble of something from a Lorentzian perspective to validate the creation rate and to analyze the behavior of field fluctuations.

\subsection{Comparison to other results}
Our results can be tested by comparing with \cite{GarciaEtxebarria:2020xsr}, where a bubble of nothing for a $T^3$ compactification in Einstein-Dilaton-Gauss-Bonnet gravity was constructed.
This construction also allows for a string-theoretic embedding.
What we referred to as ETW defect can be identified with the `inner bubble region' of \cite{GarciaEtxebarria:2020xsr}.
Naively, we would compare their results with our calculations for $d=7$.
However, the presence of the dilaton in \cite{GarciaEtxebarria:2020xsr} modifies the theory such that the resulting equations of motion are analogous to the $d=5$ theory without a dilaton. Hence, one should compare with equations derived in Sections \ref{sect_modified_Witten},\ref{sect_decay_rate}.
Using results from \cite{GarciaEtxebarria:2020xsr}, the deficit angle can be calculated to be \footnote{Here, we use the profile of the volume modulus (which is identical to the dilaton profile) at the surface where the outer bubble-region matches the inner-bubble region. More precisely, we refer to Eq. (5.87) in \cite{GarciaEtxebarria:2020xsr}.}
\begin{align}
    1-\frac{\theta_{T^3}}{2\pi}=\frac{12\pi\lambda}{(2\pi\eta)^2}\,,\label{theta_T3}
\end{align}
with $\lambda$ denoting the Gauss-Bonnet coupling and $(2\pi\eta)^3$ being the volume of the $T^3$ at the locus of the ETW defect.
The critical bubble radius $\rho_{0,T^3}$ as well as the decay exponent $B_{T^3}$ (assuming $\eta\ll R_{KK}$ where analytical results were obtained in \cite{GarciaEtxebarria:2020xsr}) are
\begin{align}
    \rho_{0,T^3} = \frac{(2\pi\eta)^2R_{KK}}{12\pi\lambda}\,\,\,,\qquad
    B_{T^3}= \pi^2M_P^2\rho_0^2\,.
\end{align}
Using \eqref{theta_T3}, one finds
\begin{align}
    \rho_{0,T^3}= \frac{R_{KK}}{1-\frac{\theta_{T^3}}{2\pi}}\,\,\,,\qquad
    B_{T^3}= \frac{\pi^2 M_P^2R_{KK}^2}{\left(1-\frac{\theta_{T^3}}{2\pi}\right)^2}\,,
\end{align}
in perfect agreement with our previous results \eqref{rho0_5d},\eqref{B_Witten_modified}.

\section{Towards a realization in type IIB string theory}\label{sect_landscape_applications}
ETW branes for Calabi-Yau or Calabi-Yau orientifold compactifications, as discussed in the previous section, could be responsible for bubble of nothing decays or bubble of something creation processes for many vacua of the string landscape. In the present section, we propose an ETW brane for type IIB compactifications on CY orientifolds with O3 planes. 
The ETW brane is constructed in a supergravity approximation of string theory and the issue of Kahler moduli stabilization is not addressed. In this sense, the level of explicitness is similar to that of GKP~\cite{Giddings:2001yu}.
We will comment on the relation to the shrinking-CY analysis of Sec.~\ref{sect_CY} at the end of Sec.~\ref{tetw}.
\subsection{The ETW brane}
\label{tetw}
The main idea is most easily explained by starting with type IIA theory on a CY manifold $Y_A$. 
The domain wall arises by considering the geometry $\mathbb{R}^4\times Y_A$, with $\mathbb{R}^4$ parameterized by $(X^0,\cdots,X^3)$, and applying a standard orientifold projection $X^3\to -X^3$. As a result, one obtains an O8 plane located at $X^3=0$ which also plays the role of an ETW brane.\footnote{
In 
the present section, we use Lorentzian geometry with a time-like ETW brane. This makes it simpler to discuss SUSY and SUSY breaking. We assume that we are allowed to return to Euclidean instanton solutions as usual, by analytic continuation in the 4d EFT.
}

Applying mirror symmetry and recalling its Strominger-Yau-Zaslow (SYZ) interpretation as T-duality of the $T^3$ fibre of the CY \cite{Strominger:1996it}, it is clear that we obtain type IIB string theory on $\mathbb{R}^4\times Y_B$ with a set of O5 planes wrapping a three-cycle of $Y_B$ and filling out three directions of $\mathbb{R}^4$. 
Clearly, this slightly more complicated ETW brane may also be constructed by directly applying an appropriate orientifold projection to $\mathbb{R}^4\times Y_B$ on the type IIB side.
We place one D5 brane on top of each O5 plane to cancel tadpoles locally.

We can go beyond plain $\mathcal{N}=2$ in the bulk by combining the orientifold projection above with a holomorphic orientifold projection acting only on $Y_B$ and introducing (for simplicity) O3 planes only. Now, away from the ETW brane, we have a type IIB ${\cal N}=1$ orientifold model of the type underlying the best understood part of the landscape.
At the locus of the ETW brane, the simultaneous existence the O5/D5 and O3/D3 systems might break SUSY completely.
However, as is argued in Appendix \ref{appendix_sect_SUSY_O3_O5}, this configuration preserves two supercharges leading to $\mathcal{N}=1$ SUSY on the 3d worldvolume of the domain wall. 

The above construction can be conveniently illustrated in the following toy-model:
Consider the space $T^6/\mathbb{Z}_2$, with the $T^6$ parameterized by three complex coordinates
\begin{align}
	{Z^i=U^i+iV^i}\,,\quad U^i\sim U^i+2\pi\,,\quad V^i\sim V^i+2\pi\,,\quad i\in\{1,2,3\}\,.\label{T6_param}
\end{align}
The $\mathbb{Z}_2$ generator acts as $(Z^i\to -Z^i,\;(-1)^{F_L}\Omega)$, with $F_L$ the left-moving fermion number and $\Omega$ the worldsheet parity. This orientifold action produces $64$ O3 planes located at $U^i=0$ or $U^i=\pi$ and $V^i=0$ or $V^i=\pi$ for $i=1,2,3$ and filling the non-compact directions.
In addition, we introduce an O5 plane generator acting by $(X^3\to -X^3$, $Z^i\to\overline{Z}^i+i\pi,\;\Omega)$.
The fixed points of this action, which determine the positions of the O5 planes, are given by
$V^i=\pm \pi/2$.
The full orientifold group is $\mathbb{Z_2}\times \mathbb{Z}_2$ containing the identity, the two aforementioned generators and their product acting by
$X^3\to -X^3$, $Z^i\to -\overline{Z}^i-i\pi$ without orientation reversal (since $\Omega^2=1$). Modding out this product generator does not induce additional singularities since the map $V^i\to V^i-\pi$ on a torus is fixed-point-free.
Table \ref{table_orientifold} summarizes the generators and Table \ref{table_orientifold_extnesion} the locations of the O-planes.
Note that the O3 and O5 planes do not intersect due to their separation along $V^i$.
We can explicitly check that this configuration does not break SUSY completely.
To show this, we first note that O$p$ planes preserve the same supercharges as D$p$ branes, such that we can equivalently think about the SUSY breaking of D3 and D5 branes. 
From Table \ref{table_orientifold_extnesion}, we can directly read off that an open string stretching between those branes is defined by four Dirichlet-Neumann boundary conditions which is known to preserve $1/4$ of the $32$ supercharges.
\begin{table}
	\centering
	\begin{tabular}{c|c|c|c|c|c|c|c|c|c|c|c}
		& $X^0$ & $X^1$ & $X^2$ & $X^3$ & $U^1$ & $V^1$ & $U^2$ & $V^2$ & $U^3$ & $V^3$ &  \\
		\hline
		$g_1$ & $X^0$ & $X^1$ & $X^2$ & $X^3$ & $-U^1$ & $-V^1$ & $-U^2$ & $-V^2$ & $-U^3$ & $-V^3$ & $\Omega (-1)^{F_L}$ \\
		\hline
		$g_2$ & $X^0$ & $X^1$ & $X^2$ & $-X^3$ & $U^1$ & $-V^1+\pi$ & $U^2$ & $-V^2+\pi$ & $U^3$ & $-V^3+\pi$ & $\Omega$\\
		\hline
		$g_1\cdot g_2$ & $X^0$ & $X^1$ & $X^2$ & $-X^3$ & $-U^1$ & $V^1-\pi$ & $-U^2$ & $V^2-\pi$ & $-U^3$ & $V^3-\pi$ & $(-1)^{F_L}$
	\end{tabular}
	\caption{Action of the two orientifold generators (of O3 and O5 planes) and of their product.}
	\label{table_orientifold}
\end{table}
\begin{table}
	\centering
	\begin{tabular}{c|c|c|c|c|c|c|c|c|c|c}
		& $X^0$ & $X^1$ & $X^2$ & $X^3$ & $U^1$ & $V^1$ & $U^2$ & $V^2$ & $U^3$ & $V^3$ \\
		\hline
		O3 & $\checkmark$ & $\checkmark$ & $\checkmark$ & $\checkmark$ & $\cross$ & $\cross$ & $\cross$ & $\cross$ & $\cross$ & $\cross$ \\
		\hline
		O5 & $\checkmark$ & $\checkmark$ & $\checkmark$ & $\cross$ & $\checkmark$ & $\cross$ & $\checkmark$ & $\cross$ & $\checkmark$ & $\cross$ \\
	\end{tabular}
	\caption{Summary of dimensions filled by O3/O5 planes (indicated with a $\checkmark$).}
	\label{table_orientifold_extnesion}
\end{table}

We expect our $T^6/\mathbb{Z}_2$ toy-model to generalize to proper CY orientifolds with O3 planes, as described at the beginning of this section. In particular, the O5 planes in type IIB are related by mirror symmetry to the O8 plane on the IIA side, which arises from the reflection $X^3\to -X^3$. Viewing $Y_B$ in terms of the SYZ fibration $T^3\to Y_B\to B_3$, with $B_3$ denoting the base, it is clear that the O5 planes are compatible with the symplectic structure of $Y_B$ (i.e. they wrap $B_3$). 
We cancel their D5 tadpole locally by placing one D5 brane on top of every O5 plane. 
One might by concerned that the O5/D5 system induces a D1 tadpole since the Chern-Simons brane action contains a term
\begin{align}
    S_{C.S.}\supset \int_{\rm O5/D5} C_2\wedge p_1(R)\,.\label{CS_coupling_5}
\end{align}
However, since the O5/D5 systems wraps a three-cycle of $Y_B$ and the first Pontryagin class $p_1(R)$ is a four-form, the above integral vanishes. Hence, our domain wall does not introduce any uncancelled tadpoles. Following this line of thought, we are lead to refine our initial statements about the O8/D8 system on the mirror-dual IIA side. Indeed, we expect the O8/D8 system to carry a non-trivial gauge bundle. To see this, note that this system wraps the entire CY. The coupling analogous to \eqref{CS_coupling_5},
\begin{align}
    S_{C.S.}\supset \int_{\rm O8/D8} C_5\wedge p_1(R)\,,\label{CS_coupling_8}
\end{align}
then induces a D4 charge, which we may cancel by an appropriately wrapped D4-brane. This brane can be taken to be resolved in the D8-stack, inducing a non-trivial gauge bundle and thereby ensuring that the IIA side is tadpole-free as well.
For completeness, we recall that the IIA side in addition has an O6/D6 system and the IIB side a mirror-dual O3/D3 system, such that the non-empty side of the ETW-brane is ${\cal N}=1$.

To be sure that the above construction forms an ETW brane of a string theory vacuum, one would need to study its backreaction more properly and address the issue of moduli stabilization.
More specifically, corrections to the Kahler moduli Kahler potential and corresponding non-perturbative superpotential effects are not considered in the bulk and hence Kahler moduli remain unstabilized.
The next step would be to include such effects and study their interplay with corresponding effects from the ETW brane. However, we are for the moment satisfied with the construction of an ETW brane at a level of explicitness similar to GKP  \cite{Giddings:2001yu}.

Let us comment in more detail on the dynamics of our effective ETW brane. At tree level this brane, viewed as an object of the 4d EFT, has vanishing tension: $T_4=0$. This is clear since we use CY geometry and cancel the tadpoles of all O5 planes involved locally. The O3 plane charge can of course only be cancelled globally, which implies non-vanishing $F_5$ field strength and some warping but does not destabilize the compact space away from the ETW brane or the brane itself. This is also consistent with the fact mentioned above (cf.~App.~\ref{appendix_sect_SUSY_O3_O5} for details) that the ETW brane preserves $\mathcal{N}=1$ SUSY in 3d.

Even in the absence of fluxes, non-perturbative effects will in general induce a non-zero superpotential, leading to an $\mathcal{N}=1$ AdS vacuum or to a runaway to infinite volume in the bulk. Since our domain wall preserves some supercharges, one might expect that it also receives corresponding corrections, giving it a tension of exactly the right magnitude to keep it flat \cite{Cvetic:1992bf,Cvetic:1996vr,Ceresole:2006iq}. A flat domain wall with critical tension $T_4=\pm T_c =\pm 2/\ell_{AdS}$ allows for a critical bubble of something (positive sign) or a critical bubble of nothing (negative sign) with infinite radius. Thus, neither of the two corresponding tunneling processes can dynamically arise.

We note, however, that the above relies on an analysis performed for domain walls in a 4d theory with smoothly varying superpotential\cite{Cvetic:1992bf,Cvetic:1996vr,Ceresole:2006iq}. It is not clear to us whether the conclusion concerning the critical tension can really be carried over to our case of a (generically singular) ETW brane. In particular, we feel that we can not firmly exclude the possibility of supersymmetric but non-stationary ETW branes with $AdS_3$ worldvolume geometry. For instance, it was shown in \cite{Bagger:2002rw} that 5-dimensional supergravity does admit supersymmetric domain walls with ``arbitrary'' tension. In 4d, one might expect that the ordinary BPS bound for field theory domain walls will carry over to ETW branes, although this is not immediately compatible with the following holographic argument: Find an AdS${}_3$ supergravity theory containing a large-$N$ CFT in the matter sector. This CFT can be holographically interpreted as an additional spatial dimension, implying the existence of an AdS${}_4$ `bulk space'. The AdS${}_3$ we started from now plays the role of the boundary of our 4d theory. Being AdS, this boundary is non-stationary and corresponds to a `detuned' ETW brane. 

In any event, even if such `detuned' ETW branes should be relevant in our setting, both types of bubble-of-anything  processes are still excluded since the tension lies in the range $-T_c\leq T_4\leq T_c$.
Thus, for bubbles of either type to arise, SUSY must be broken in the bulk or by the ETW brane. We will now discuss both options.

First, let us focus on bulk SUSY breaking. This may arise if our ${\cal N}=1$ vacuum is modified by further effects, such as fluxes and/or anti-D3 brane uplifts etc., leading to a metastable SUSY breaking minimum. If fluxes are present, they would first have to be removed by an appropriate domain wall consisting of wrapped D5/NS5 branes, which are extended in parallel to our O5/D5 ETW brane. We will from now on view both together as our effective ETW brane.

The bulk SUSY breaking effects will clearly modify both the vacuum energy and the tension of the domain wall. Most naively, one expects that SUSY breaking increases the decay probability, i.e.~that bubbles of nothing become possible similar to \cite{GarciaEtxebarria:2020xsr}. 
However, especially if the tension has been positive before SUSY breaking, it is also plausible that a bubble of something arises.
We expect the correction to $T_4$ to be set by a combination of the KK scale and the SUSY-breaking scale.

If, instead of the supersymmetric ETW brane constructed above, we manage to find an ETW brane that does not preserve any supercharges, we expect the tension to be larger such that bubbles of something may become allowed. For example, one may imagine that the D5 branes which are part of our ETW construction are wrapped on cycles that are distinct from the special lagrangian 3-cycles of the O5-planes. Whether this is possible while still cancelling the O5-tadpole and ensuring a sufficiently metastable or long-lived ETW brane is an open question.
It appears likely that the small-$\eta$ ETW defects speculated to exist in Sec. \ref{sect_generalized_Witten} are SUSY-breaking and hence provide an example of the SUSY-breaking ETW branes just discussed.
Generally, since non-supersymmetric defects have been predicted by the cobordism conjecture \cite{McNamara:2019rup,Dierigl:2022reg,Debray:2023yrs}, there likely exist ETW branes suitable for bubble of something creation.
However, it is a general challenge to explicitly construct metastable objects with large tension and it could be a similarly hard task as finding metastable de Sitter vacua in string theory.\footnote{For example, one expects a 9d non-SUSY positive-tension ETW brane for the 10d Minkowski type-IIB vacuum to exist. Its tension presumably depends on the dilaton, dynamically driving the latter to infinity. It would be interesting to study whether bubble of something processes are possible in settings of this type.}

For simplicity, we so far considered CY orientifolds with O3 planes only. 
However, nothing prevents us from also allowing for O7 planes wrapped on holomorphic cycles on the CY. The resulting 4d vacua still have $\mathcal{N}=1$ SUSY and, as argued in App.~\ref{appendix_sect_SUSY_O3_O5}, the O5/D5 ETW brane still preserves two supercharges. However, the O7 and the O5 plane now generically intersect.

Note that the $T^6$ example above also allows for the introduction of O7 planes: Consider a compactification of type IIB string theory on $T^2/\mathbb{Z}_2\times T^4$ containing four O7 planes.
Using again the parametrization \eqref{T6_param} with $\mathbb{Z}_2$ action $(Z^1\to-Z^1,\;(-1)^{F_L}\Omega )$, the O7 planes are located at points where $U^1,V^1$ take values $0$ or $\pi$.
Introducing again an O5 plane generator compatible with the symplectic structure of the torus, $(X^3\to-X^3,Z^i\to\overline{Z}^i+i\pi,\;\Omega)$, one sees that the O5 planes are located at $U^i=\pm \pi/2,\, V^i=\pm \pi/2$.
Hence, the O5s and O7s do not intersect in this geometry.

However, it is questionable whether this property can also arise in proper CY compactifications. Here, in analogy to the torus example, the O7 planes are compatible with the holomorphic structure of the CY and the O5 planes with its symplectic structure.
The non-trivial fibration of the SYZ-torus over the base $B_3$ may make it unavoidable for the O5s and O7s to intersect.

Finally, we want to comment on the relation to the possibly more general type of CY-based ETW brane of Sec.~\ref{sect_CY}. There, the CY was assumed to shrink to some radius $\eta$ before the 4d world terminates in a suitable defect. By contrast, in the present section we analysed an ETW brane which terminated the 4d theory without the need of a previous shrinking of the CY. Once SUSY is broken (either in the bullk or on the brane) this solution has corrections and it is conceivable that, when these are included, our present solution is dynamically driven to a solution of the type discussed in Sec.~\ref{sect_CY}. Put differently, we may think of the presently discussed ETW brane as of a special case where $\eta \approx R_{KK}$. Corrections may drive this value of $\eta$ to other, e.g. $\order{1}$, values.

\subsection{Creation/decay rates}
We now want to estimate the creation/decay rates of bubbles of anything involving the ETW brane constructed above.
We assume $\eta\approx R_{KK}$, as discussed before and rely on our calculations for Minkowski vacua from Sec.~\ref{sect_generalized_Witten}. These remain approximately valid for de Sitter (dS) or AdS models as long as  $\ell_{dS/AdS}\gg M_P^2/|T_4|$, which defines the class of vacua we are considering in this section.
Clearly, vacua with cosmological constant tuned close to zero most easily fulfill this condition which is why we focus on vacua of this type.

Let us now comment on the scales setting the magnitude of $T_4$. 
The non-vanishing tension is a combination of two effects: The O3/D3 or O7/D7 system present in the bulk leading to $\mathcal{N}=1$ SUSY in 4d and the additional SUSY breaking effects either in the bulk or on the boundary.
Before SUSY breaking, the tension is generically of order $|T_4|\sim T_c\simeq M_P^2/\ell_{AdS}$. We want to express this in terms of the physical KK radius $R_{KK}/M_{10}$, with $M_{10}$ denoting the 10d Planck mass. Hence we introduce a parameter $\epsilon$ measuring the separation of scales between the KK- and the AdS curvature length
\begin{align}
    R_{KK}/M_{10}=\epsilon\ell_{AdS}\,.
\end{align}
We are at this point agnostic concerning possible constraints on realizing $\epsilon\ll 1$\cite{Gautason:2015tig, Gautason:2018gln, Lust:2019zwm}.

Eventually, the tension before SUSY breaking can be expressed as
\begin{align}
    |T_4|\sim \epsilon\frac{M_P^2}{R_{KK}/M_{10}}\sim \epsilon\frac{M_P^3}{R_{KK}^4}\,.\label{T_4_epsilon}
\end{align}
We expect that corrections from the uplifting mechanism will not invalidate this estimate. Thus we will continue to work with the ETW brane tension \eqref{T_4_epsilon}, with both signs for $T_4$ being allowed.

\paragraph{Bubbles of something:}
As discussed before, a tension $T_4>0$ suitable for a bubble of something can arise due to SUSY breaking either in the bulk or on the ETW brane. 
Since we assumed $\eta\approx R_{KK}$, the volume modulus stays approximately constant away from the ETW brane and we may use our estimate \eqref{B_BoS} for the creation rate:
\begin{align}
    B=-\frac{8\pi^2M_P^6}{T_4^2}\sim -\frac{R_{KK}^8}{\epsilon^2}\,.\label{B_BoS_epsilon}
\end{align}
If the vacuum under consideration is dS, it would be interesting to compare the above result with those of the Hartle-Hawking no-boundary and the Linde/Vilenkin tunneling proposals.
As discussed before in Section \ref{sect_bubble_of_something}, it is conceptually unclear how the different creation rates should be compared to each other, but we can nevertheless make a few observations:

The creation exponents for the no-boundary and the tunneling proposals are given by
\begin{align}
    B=\mp 8\pi^2M_P^2\ell_{dS}^2\,,\label{B_HaHa_LV}
\end{align}
with the upper sign corresponding to the Hartle-Hawking and the lower sign to the Linde/Vilenkin process. Clearly, the sign of $B$ in \eqref{B_BoS_epsilon} matches that of Hartle-Hawking, which is a straightforward consequence of our euclidean approach. It would be interesting to investigate whether a Linde/Vilenkin-type analysis of the bubble-of-something creation process can be implemented and shown to produce a positive exponent.

Assuming that \eqref{B_BoS_epsilon} may be directly compared to the Hartle-Hawking version of \eqref{B_HaHa_LV}, one can use our initial assumption $\ell_{dS}\gg M_P^2/|T_4|$ to conclude that the Hartle-Hawking process is more likely to occur than a bubble-of-something process. It is straightforward to check this remains true after a more general analysis, relaxing the above assumption and including the effects of dS geometry.

\paragraph{Bubbles of nothing:}
SUSY breaking in the bulk can lead to the existence of a bubble of nothing with $T_4<0$.
The scales setting the tension are already discussed above.
To estimate the decay rate, we approximate the instanton geometry away from the ETW brane by flat space, in analogy to Sec. \ref{sect_bubble_of_something}.
The decay exponent follows from a calculation similar to \eqref{K4_BoS_naive}-\eqref{B_BoS}, but its sign is reversed due to the extrinsic curvature now being negative:
\begin{align}
    B=\frac{8\pi^2M_P^6}{T_4^2}\sim \frac{R_{KK}^8}{\epsilon^2}\,.\label{B_O5_ETW_brane_nothing}
\end{align}

The rate following from \eqref{B_O5_ETW_brane_nothing} should be compared to the rates of other decay channels, such as tunneling to decompactification, flux transitions or the Kachru-Pearson-Verlinde (KPV) \cite{Kachru:2002gs} process.
For traditional dS constructions in string theory \cite{Kachru:2003aw,Balasubramanian:2005zx,Conlon:2005ki}, the decay exponents of tunneling to decompactification and flux transitions are, up to a few orders of magnitude\footnote{For example, in KKLT models the tunneling exponent can be suppressed by factors of the volume.}, close to \cite{Kachru:2003aw,Westphal:2007xd,deAlwis:2013gka}
\begin{align}
    B\sim M_P^2 \ell_{dS}^2\,,\label{B_ldS}
\end{align}
with $\ell_{dS}$ being the de Sitter length.
The above becomes more exact the closer the dS vacuum energy approaches zero. The value of $B$ in \eqref{B_ldS} is extremely large, of the order of magnitude needed to saturate the de Sitter lifetime.
We thus expect bubble of nothing decays to occur faster. This also follows explicitly since, by comparing to \eqref{B_O5_ETW_brane_nothing}, we immediately see that if $\ell_{dS}^2T_4^2>M_P^4$ bubble of nothing decays are faster. But this condition was already assumed to hold at the beginning of the section, such that it is automatically satisfied for all vacua we are considering at the moment.

If the vacuum contains a Klebanov-Strassler throat \cite{Klebanov:2000hb} with $\overline{D3}$ branes as an uplifting mechanism, the KPV decay process \cite{Kachru:2002gs} might occur. 
Generically, this decay happens in the field-theoretic regime and the decay exponent is given by \cite{Kachru:2002gs,Freivogel:2008wm} 
\begin{align}
    B_{KPV}= \frac{27\,(0.932)^6 \,M^6\,g_s}{2048\pi\, (N_{\overline{D3}})^3}\,.\label{B_KPV}
\end{align}
Here, $M$ denotes the flux number on the A-cycle, $N_{\overline{D3}}$ the number of anti-D3 branes and $g_s$ is the string coupling.
We see that this $B$ is not parametrically large but only enhanced by a largish integer $M$ needed to avoid a classical instability. The KPV estimate $g_sM^2\gtrsim 12$ is under debate\cite{Bena:2018fqc, Blumenhagen:2019qcg, Lust:2022xoq}. This instability issue endangers the practicality of the KPV uplift, see e.g.~\cite{Carta:2019rhx, Gao:2020xqh, Junghans:2022exo, Gao:2022fdi}. In particular, it has been argued that practical applications must consider KPV with NS5-brane curvature corrections, which change the potential barrier significantly \cite{Hebecker:2022zme, Schreyer:2022len}. 
Remembering that we estimated $|T_4|$ to be parametrically bounded from above by $\epsilon R_{KK}^6$, one may expect our $B\sim R_{KK}^8/\epsilon^2$ to be larger than that of KPV. However, this story is still evolving.

Finally, there remains the possibility of the existence of exotic domain walls with $|\theta|\gg 1$, which can lead to a significant enhancement of the decay rate to nothing. 

\section{Bubbles of anything -- further remarks}\label{sect_comments}
In this section, we collect further comments on bubbles of anything and discuss how the creation/decay rates can be computed more precisely in the future.

\subsection{Nothing as AdS with infinitely negative vacuum energy} 
\begin{figure}[h]
	\centering
 \includegraphics[width=0.65\linewidth]{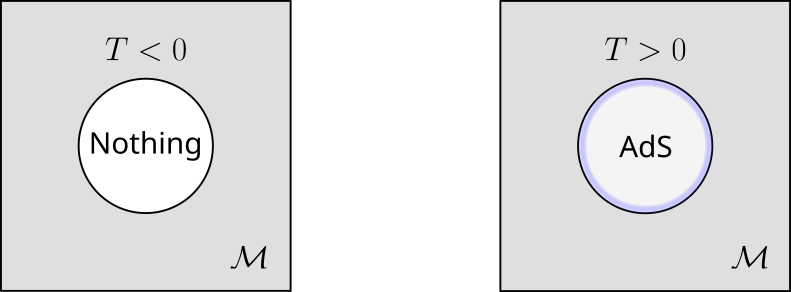}
	\caption{Instantons for a bubble of nothing decay (left) and a decay to AdS (right) of Minkowski space. In the second case, the ball of AdS glued into the flat space ${\cal M}$ has most of its contribution to the action localized near the boundary, here illustrated by the blue-shaded region.}%
	\label{fig_Instanton_BoN}
\end{figure}

\begin{figure}[h]
	\centering
 \def\svgwidth{0.45\linewidth}
 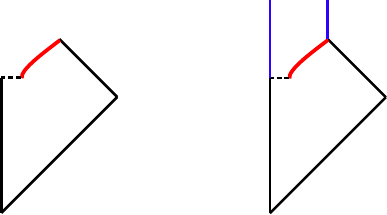
	\caption{We display the Penrose diagram for a Minkowski bubble of nothing (left) and a Minkowski to AdS decay (right). On the left, the red line denotes the ETW brane. On the right, the blue lines characterize the AdS region and the red line represents the domain wall. In both diagrams, the dashed line denotes the quantum process of the critical bubble emerging. The ETW brane on the left requires a negative tension while the domain wall on the right has positive tension.}%
	\label{fig_penrose_nothing}
\end{figure}

\begin{figure}[h]
	\centering
 \def\svgwidth{0.45\linewidth}
 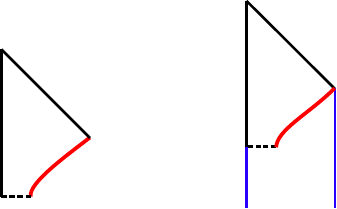
	\caption{The Penrose diagrams for a Minkowski bubble of something (left) and a hypothetical AdS to Minkowski transition (right) are displayed. The blue region denotes the AdS region and the red lines correspond to the ETW brane (left) and domain wall (right) respectively.}%
	\label{fig_penrose_something}
\end{figure}

In \cite{Brown:2011gt}, it was suggested that `nothing' can be modelled as an AdS space in the limit of infinitely negative vacuum energy or, equivalently, infinitely small AdS length $\ell_{AdS}$. Indeed, a ball of AdS of radius $R\gg\ell_{AdS}$ has most of its negative energy localized near the boundary, cf.~r.h.s.~of Fig.~\ref{fig_Instanton_BoN}.
Hence a brane separating Minkowski or de Sitter space\footnote{
For simplicity, all formulae and the more detailed discussions below are performed only in the $\ell_{dS}\to \infty$ limit, which is completely sufficient for the qualitative purposes of the present section.
} 
and an AdS region (with $\ell_{dS}\gg R\gg \ell_{AdS}$) may be seen as an effective ETW brane. Its tension receives a correction from the `thin', negative-energy AdS region on one side of the brane. As a result, tunneling between Minkowski/dS and AdS 
may be a good model for bubble of anything processes. In Fig. \ref{fig_penrose_nothing}, the Penrose diagrams for the decay of 
Minkowski space through a bubble of nothing and its decay to AdS are shown. The diagrams for the (hypothetical) inverse processes are given in Fig. \ref{fig_penrose_something}. Using the analogy between the l.h.~and the r.h.~sides of Figs.~\ref{fig_Instanton_BoN}-\ref{fig_penrose_something} and assuming that up-tunneling from AdS is impossible, one may then claim that the corresponding bubbles of something are forbidden \cite{Brown:2011gt}. We want to argue against this conclusion.

To be more precise (cf.~App.~\ref{appendix_sect_AdS_negative_tension} for supporting calculations), consider first tunneling from Minkowski/dS to AdS. As explained above, the instanton is flat space or a large sphere (the euclidean version of the decaying dS), with a small ball of AdS glued in. It is bounded by a brane with positive tension $T$, cf.~Fig. \ref{fig_Instanton_BoN}. When evaluating the action of this instanton configuration, one notices that the AdS on the inner side of the spherical, fundamental brane with tension $T$ contributes an amount $-4\pi^2R^3/\ell_{AdS}$, with $R$ being the radius of the bubble. 
The cubic scaling in $R$ is similar to the contribution from the fundamental brane, which contributes $2\pi^2TR^3$. This can be understood at the intuitive level by noticing that the AdS has most of its volume localized near the boundary.
As a result, the AdS and the brane taken together behave like an ETW brane with total tension $T_{eff}=T-2/\ell_{AdS}$.
As shown in App.~\ref{appendix_sect_AdS_negative_tension}, a nonzero transition rate requires $T_{eff}<0$.
One then takes the limit $\ell_{AdS}\to 0$, $T\to \infty$ while keeping $T_{eff}$ fixed, ensuring that the critical bubble radius stays finite. Given this renormalization procedure, one finds that the Brown-Dahlen model \cite{Brown:2011gt} perfectly describes a decay of dS/Minkowski to nothing.

The suggestion of \cite{Brown:2011gt} is that the exponent for up-tunneling from AdS to dS should be calculated using the same instanton, but this time subtracting not a full dS sphere but rather an infinite AdS as background. This clearly leads to a divergence and hence to a vanishing tunneling rate. However, this divergence comes from the infinite AdS volume. This is in conflict with our initial motivation for using AdS as a model for `nothing', which relied on the claim that only a thin slice of AdS contributes to the action. Moreover, this whole approach is not consistent with the standard Coleman-De Luccia procedure \cite{Coleman:1980aw}, where the instanton is constructed by gluing a finite patch of the new vacuum into the background vacuum.

In our opinion, a more appropriate approach to a potential transition AdS$\,\to\,$Minkowski/dS is to construct the instanton following \cite{Coleman:1980aw}: One glues a small ball of Minkowski or dS into infinite AdS. The subtracted background is also AdS and the divergence from the infinite AdS cancels between instanton and background. Only a thin slice of the background AdS on the inner side of the domain wall of radius $R$ contributes as $+4\pi^2R^3/\ell_{AdS}$ to the action. This time the effective ETW brane tension reads $T_{eff}=T+2/\ell_{AdS}$, with the fundamental tension $T$ necessarily being negative. Together with the condition $T_{eff}>0$, this ensures the existence of a solution to Einsteins equations. 
However, negative-tension branes suffer from instabilities since it is energetically favorable to increase their volume. As a result, it is widely believed that they can not mediate vacuum transitions. This would imply that up-tunneling from AdS is not a good model for bubbles of something. However, we can not completely exclude the existence of negative-tension domain walls, especially in the case where one has strongly curved AdS on one of its sides together with a suitably low UV cutoff. If such branes can indeed be constructed, one could complete the analysis above, performing the required `renormalization': Taking $\ell_{AdS}$ to zero, $T$ to $-\infty$ and keeping $T_{eff}$ positive and finite, one would find an AdS-based model for bubbles of something.

Our conclusion is that, at the current level of knowledge, `nothing' can not be universally viewed as AdS in the limit $\ell_{AdS}\to 0$: 
Although the model of \cite{Brown:2011gt} can describe bubble of nothing decays perfectly, it fails to describe bubble of something creation events. The reason is that in the second case negative-tension domain walls are required and these are in general unstable. While it may be worthwhile to try and overcome this obstacle, the implications for us are in any case clear: Either any such efforts fail, then AdS is not a universal model for `nothing'. Or they succeed, then bubbles of something can also be recovered from this AdS-based perspective. In both cases we conclude that the Brown-Dahlen approach does {\it not} exclude bubble-of-something creation events.

A different way of relating deep AdS and `nothing' is discussed in \cite{Cespedes:2023jdk}.

\subsection{Refining the decay/creation rate estimates}
Our calculations from Secs.~\ref{sect_generalized_Witten} and \ref{sect_landscape_applications} provide only leading-order estimates for the decay/creation rates. While we have to leave a more precise treatment to future work, we want to at least comment on the necessary improvements.

\subsubsection{Non-trivial potentials}
In the present paper, we have focused on compactifications to Minkowski space with vanishing potential for the volume modulus.
Clearly, in any realistic compactification, a stabilizing potential must exist. 
Nevertheless, the general strategy for finding bubble of nothing/something solutions stays similar to what is discussed in Secs. \ref{sect_dynamics} and \ref{sect_generalized_Witten}. 
However, it now becomes a non-trivial task to solve the equations of motion \eqref{phi_eom}, \eqref{f_eom}, subject to the boundary conditions \eqref{eta_def}, \eqref{theta_def_CY}.
The calculation of the extrinsic curvature, its relation to the tension $T_4$ \eqref{K_T_relation} and its contribution to the tunneling action are unchanged.
In addition to the boundary contribution to $B$ \eqref{decay_rate_general}, there will now also be a bulk term.
For both the instanton and the vacuum solution, this contribution to the action reads \eqref{S_int_on_shell}
\begin{align}
    S_{bulk} = -2\pi^2\int f^3V(\phi)dr\,.\label{B_contribution_bulk}
\end{align}
This has to be evaluated using the profile $\phi(r)$ obtained by solving the equations of motion.
For some analytic and numerical estimates in 5d bubble of nothing models see \cite{Draper:2021qtc,Draper:2021ujg,Draper:2023ulp}.

Allowing for non-trivial potentials includes the generalization to situations where the decaying/created vacuum is dS or AdS. The resulting modifications are straightforward. Moreover, if $\ell_{dS/AdS}$ is large compared to the critical bubble radius, we expect only small corrections relative to the results derived in the previous sections.
If the considered vacuum is dS, one can not distinguish between a bubble of nothing and a bubble of something instanton. Any solution can be interpreted in both ways, which is a result of euclidean dS space being compact.
This is analogous to Coleman-de Luccia instantons for dS $\to$ dS tunneling where the same instanton can be viewed as an up- or down-tunneling event and only the difference in the vacuum contribution to the decay exponent \eqref{decay_rate_general} leads to a suppressed up-tunneling rate.

In models where $\phi$ stays approximately constant away from an ETW brane with tension $T_4$, cf. Sec. \ref{sect_bubble_of_something}, we can directly calculate the dS and AdS decay/creation exponents.
\paragraph{de Sitter}
For the dS case, similar calculations in a slightly different setting have appeared in \cite{Bousso:1998pk}.
The ETW brane sits at some radius $f_\eta\leq\ell_{dS}$ of the euclidean dS sphere. The extrinsic curvature and its relation to the tension reads
\begin{align}
    \frac{3}{2}T_4=M_P^2\mathcal{K}_4=\pm \frac{3M_P^2}{f_{\eta}}\sqrt{1-\frac{f_{\eta}^2}{\ell_{dS}^2}}\,,\label{T4_R_rel}
\end{align}
where the positive sign corresponds to a dS patch of size less than a hemisphere and the negative sign to a dS patch of size larger than a hemisphere.
The instanton action and thus the creation exponent is given by
\begin{align}
    S_{instanton}=4\pi^2M_P^2\ell_{dS}^2\left(\sqrt{\frac{T_4^2\ell_{dS}^2}{T_4^2\ell_{dS}^2+4M_P^4}}-1\right)
    \label{si1}
\end{align}
if the tension is positive and 
\begin{align}
    S_{instanton}=-8\pi^2M_P^2\ell_{dS}^2-4\pi^2M_P^2\ell_{dS}^2\left(\sqrt{\frac{T_4^2\ell_{dS}^2}{T_4^2\ell_{dS}^2+4M_P^4}}-1\right)
    \label{si2}
\end{align}
if the tension is negative.
To obtain the decay exponent, one needs to subtract the vacuum action $S_{vaccum}=-8\pi^2M_P^2\ell_{dS}^2$ from $S_{instanton}$.

\begin{figure}[h]
	\centering
 \def\svgwidth{0.7\linewidth}
 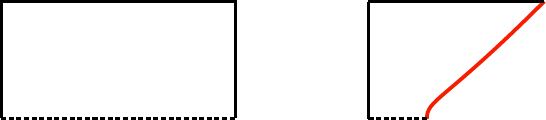
	\caption{Penrose diagrams of a Hartle-Hawking or Linde/Vilenkin process (left) and of a bubble of something leading to dS (right). The first is simply half a dS space, while the latter contains an ETW brane boundary (red line). The dashed lines indicate the inital universe created from nothing via a quantum process.}%
	\label{fig_penrose_LV}
\end{figure}

Let us emphasize a key differences between a Hartle-Hawking \cite{Hartle:1983ai} or Linde/Vilenkin  \cite{Linde:1983mx, Vilenkin:1984wp} process and a bubble of something event. The corresponding Penrose diagrams are shown in Fig. 
\ref{fig_penrose_LV}. While a Hartle-Hawking or Linke/Vilenkin process nucleates a closed universe (a sphere) with its initial size necessarily being the dS radius, a bubble of something produces a ball bounded by an ETW brane. Hence, the critical bubble radius depends on the brane tension and can in particular be much smaller than the dS radius.

\paragraph{Anti de Sitter}
Here, similar to Minkowski space, a bubble of nothing instanton is a non-compact patch of AdS with a hole of radius $f_\eta$ meanwhile a bubble of something instanton is a compact patch of AdS of radius $f_\eta$.
The former requires a negative ETW brane tension and the latter a positive one.
The relation between the extrinsic curvature and the tension reads
\begin{align}
    \frac{3}{2}T_4=M_P^2\mathcal{K}_4=\pm \frac{3M_P^2}{f_\eta}\sqrt{1+\frac{f_\eta^2}{\ell_{AdS}^2}}\,,
\end{align}
with the upper/lower sign appropriate for a bubble of something/nothing.
One immediately sees that $|T_4|>2M_P^2/\ell_{AdS}$ is a necessary requirement to have any bubble.
The creation/decay exponents are then given by
\begin{align}
    B_{creation}&=-4\pi^2M_P^2\ell_{AdS}^2\left(\sqrt{\frac{T_4^2\ell_{AdS}^2}{T_4^2\ell_{AdS}^2-4M_P^4}}-1\right)\,,\\
    B_{decay}&=4\pi^2M_P^2\ell_{AdS}^2\left(\sqrt{\frac{T_4^2\ell_{AdS}^2}{T_4^2\ell_{AdS}^2-4M_P^4}}-1\right)\,.
\end{align}

Using the above results, one can verify that holding fix a tension $T_4$, bubbles of nothing/something become more probable if the vacuum energy increases/decreases.

\subsubsection{Fluxes}
If the CY compactification involves fluxes, one generally requires charged ETW branes. 
Such ETW branes may be constructed by noting that brane flux transitions are possible \cite{Kachru:2002gs}, realized by wrapped branes forming a domain wall in the non-compact directions. Bubble of nothing decays can then occur as a `double process', where the removal of the flux by a brane is followed by a decay to nothing along the lines of Secs.~\ref{sect_generalized_Witten} and \ref{sect_landscape_applications}.
For a decay rate estimate in toy models, see \cite{Draper:2023ulp}. We may think of our desired charged domain wall as being realized if the flux-removing brane is moved on top of the ETW brane.
As already noted in Sec. \ref{sect_CY}, our formalism of Sec. \ref{sect_generalized_Witten} still applies with appropriate values of $\theta$ and $\eta$, which of course first have to be derived.

\subsubsection{Other fields}
In an exhaustive treatment of bubbles of anything, all fields, such as moduli fields, gauge fields and the dilaton have to be taken into account.
If those fields are only weakly stabilized, they can contribute significantly to the decay rate, as can be seen from \cite{GarciaEtxebarria:2020xsr}.
Here, in the outermost layer, two fields are dynamical, namely the volume modulus and the dilaton. 
Both fields obtain the same profile such that the equations of motion change to a form that resemble the ones from the Witten bubble of nothing analyzed in Sec. \ref{sect_modified_Witten}.
In the scenario of \cite{GarciaEtxebarria:2020xsr}, the presence of the dilaton increases the decay rate when comparing to the $d=7$ estimate obtained in App.~\ref{appendix_sect_general_dimension}.
One can also imagine situations where additional fields lead to a decrease of the decay rate.

\subsubsection{Higher-order corrections}
The supergravity action obtained as the low-energy limit of string theory receives corrections at higher orders in the $\alpha'$ expansion. 
Such corrections typically become important at small distance scales which in principle can induce additional effects in the region where the internal manifold has shrunk to small sizes.
It would be interesting to investigate the consequences of the higher-order corrections on the instanton geometry and the decay/creation rates.

\section{Conclusions}
Our main results in the present paper concern the theoretical development and implementation of `bubble of anything' processes in the string theory landscape, making them potentially relevant to cosmology.

First, we formulated a 4d EFT description for Witten's bubble of nothing based on $S^1$ compactifications.  In doing so, we also allowed for a defect with variable tension, sitting at the point where the $S^1$ disappears. Crucially, this defect has non-zero size $\eta$ and induces a deficit angle $\theta$. This deficit angle measures the tension of the defect and contributes to the effective 4d tension $T_4$ of the boundary. In such a more general setting, both bubble of nothing and bubble of something instantons exist. The latter require a non-trivial potential for the  scalar governing the $S^1$ radius as well as a positive effective ETW brane tension $T_4$.

We have demonstrated that the results above may be generalized to Calabi-Yau orientifold compactifications, again with the assumption that the compact manifold shrinks to zero size at the boundary of spacetime. Since the microscopic structure of Calabi-Yau (CY) bordisms to nothing is generally unknown, the introduction of an appropriate defect of size $\eta$ and with a (suitably generalized) deficit angle $\theta$ is in this case mandatory. While it may be natural to assume the defect size to be of higher-dimensional Planck or string scale, it can in principle also be much larger. We provide general formulae for the actions of Calabi-Yau bubbles of anything of this type, which turn out to be parametrically larger than for the Witten bubble. These actions allow for estimates of the decay or creation rates of corresponding 4d vacua.

While we feel that it is well-motivated by the Cobordism Conjecture, the ad-hoc introduction of a defect realizing the bordism of a Calabi-Yau orientifold to nothing is certainly unsatisfactory. We therefore continued our analysis by also proposing a construction of an ETW brane for type IIB Calabi-Yau orientifold models with O3/O7 planes. 
Our construction employs an O5 orientifolding of the type IIB compactification, involving a reflection in one of the non-compact spatial dimensions.
The O5/D5 system wraps the bubble surface and a three cycle of the CY. The resulting ETW brane preserves two supercharges, i.e.~$\mathcal{N}=1$ supersymmetry in 3d.
Naively, we then expect such an ETW brane to be either stationary or contracting, allowing neither for a bubble of nothing nor of something.
However, once we include moduli stabilization and SUSY breaking in the bulk, as well as (possibly) additional SUSY breaking on the boundary, this will certainly change.
One may generically expect a combination of the KK- and SUSY breaking scale to set the brane tension. In case a bubble of nothing is allowed, one can then compare our decay rates (with decay exponent \eqref{B_O5_ETW_brane_nothing}) with the rates of traditional vacuum decay processes, e.g. the decay to decompactification, flux transitions or the Kachru-Pearson-Verlinde (KPV) decay process.
It turns out that only the field-theoretic decay of KPV is likely to be faster.

For bubbles of something, the nucleation exponent is given in \eqref{si1}, \eqref{si2} and a naive estimate for our setting in \eqref{B_BoS_epsilon}. Bubbles of something may be viewed as an alternative to the Hartle-Hawking no-boundary creation process. The form of the nucleation exponents differs significantly between these two vacuum creation processes: In one case, the key input are the ETW brane tension and 4d vacuum energy, in the other case it is only the 4d vacuum energy. One may ask whether the relative importance of the two creation processes for the landscape can be quantified by directly comparing the respective nucleation exponents.

It would be desirable to analyse the effects of the dilaton, fluxes, vacuum energy and non-perturbative effects in more detail.
We also leave a careful study of SUSY-breaking and supersymmetric corrections 
to the domain wall tension to future work.  Other open questions are whether bubbles of something can be interpreted using the Linde-Vilenkin approach to vacuum creation\cite{Linde:1983mx,Vilenkin:1984wp} and whether they can be studied in a Lorentzian approach\cite{Feldbrugge:2017kzv}.

\subsection*{Acknowledgements}
We are very grateful for useful discussions with Giuseppe Dibitetto, Luca Martucci, Jacob McNamara, Miguel Montero, Simon Schreyer, Irene Valenzuela, Gerben Venken and Timo Weigand. This work was supported by Deutsche Forschungsgemeinschaft (DFG, German Research Foundation) under Germany’s Excellence Strategy EXC 2181/1 - 390900948 (the Heidelberg STRUCTURES Excellence Cluster). 

\appendix

\section{Dimensional reduction}\label{appenndix_sect_dimensional_reduction}
In the following, we present the dimensional reduction from $d$-dimensions to four dimensions starting from the metric ansatz \eqref{d_dim_geometry}
\begin{align}
\begin{split}
    ds^2&=e^{2\alpha \phi(r)}\left(dr^2+f(r)^2d\Omega_3^2\right)+e^{2\beta \phi(r)}ds_n^2\\
    &=e^{2\alpha \phi}\left(ds_4^2+e^{2(\beta-\alpha)\phi}ds_n^2\right)=e^{2\alpha\phi}d\tilde{s}^2\,.\label{appendix_metric_alpha_beta}
    \end{split}
\end{align}
We set $M_d=1$ and normalize 
$ds_n^2$ (and hence the compact-space metric) by
\begin{align}
    \int \sqrt{g_n}d^nx =1\,,
\end{align}
such that the 4d Planck mass is unity as well.
\paragraph{Einstein-Hilbert action:}
When dimensionally reducing the $d$-dimensional Einstein-Hilbert action, it is convenient to employ the metric $d\Tilde{s}^2$ introduced in \eqref{appendix_metric_alpha_beta}.
Using the transformation properties of the Ricci scalar under Weyl rescalings, the action takes the form
\begin{align}
    S&=-\frac{1}{2}\int \sqrt{g_n}\sqrt{g_4}e^{d\alpha\phi}e^{n(\beta-\alpha)\phi}e^{-2\alpha \phi}\left(\tilde{\mathcal{R}}-2(d-1)\alpha\Tilde{\square}\phi-(d-2)(d-1)\alpha^2(\partial\phi)^2\right)\\
    &=-\frac{1}{2}\int\sqrt{g_4}e^{[2\alpha+n\beta]\phi}\left(\Tilde{\mathcal{R}}-2(d-1)\alpha \square_4 \phi -\left[(d-2)(d-1)\alpha^2 +2(d-1)n(\beta-\alpha)\alpha \right](\partial\phi)^2\right)\,,\label{appendix_action1}
\end{align}
where $\tilde{\square}_4$ is the Laplace operator associated with $d\Tilde{s}_4^2$ and $\square_4$ is the Laplace operator associated with $ds_4^2$.
Further, we have used
\begin{align}
    \Tilde{\square}\phi=\frac{1}{f^3e^{n(\beta-\alpha)\phi}}\partial_r(f^3e^{n(\beta-\alpha)\phi}\phi')=\frac{1}{f^3}\partial_r(f^3\phi')+n(\beta-\alpha)(\partial\phi)^2=\square_4\phi+n(\beta-\alpha)(\partial\phi)^2\,.
\end{align}
We proceed by calculating $\Tilde{\mathcal{R}}$, the Ricci scalar of the metric $d\Tilde{s}^2$.
To do this, we first write the metric in the form
\begin{align}
    d\Tilde{s}^2=e^{2(\beta-\alpha)\phi}\left(dz^2+g(r)^2d\Omega_3^2+ds_n^2\right)\,,
\end{align}
where we defined $e^{(\beta-\alpha)\phi}dz=dr$ and $g(r)e^{(\beta-\alpha)\phi(r)}=f(r)$.
Using twice the transformation properties of the Ricci scalar under conformal transformations one arrives at the result
\begin{align}
    \Tilde{\mathcal{R}}=\mathcal{R}_4-2n(\beta-\alpha)\square_4\phi-n(n+1)(\beta-\alpha)^2(\partial\phi)^2+\mathcal{R}_ne^{-2(\beta-\alpha)\phi}\,,
\end{align}
with $\mathcal{R}_4$ denoting the Ricci scalar of the metric $ds_4^2$ and $\mathcal{R}_n$ the Ricci scalar of the metric $ds_n^2$. For completeness, we note that
\begin{align}
    \mathcal{R}_4=6\left(\frac{1}{f^2}-\frac{f'^2}{f^2}-\frac{f''}{f}\right)\,.
\end{align}
From \eqref{appendix_action1} one observes that setting
\begin{align}
    \beta=-2\alpha/n=-2\alpha/(d-4)\label{appendix_alpha_beta_rel}
\end{align}
leads to a 4d Einstein frame action (recall that $d=n+4$)
\begin{align}
    S=\int\sqrt{g_4}\left(-\frac{1}{2}\mathcal{R}_4+\alpha \square_4\phi+\frac{2(n+2)}{n}\alpha^2\frac{(\partial\phi)^2}{2}-\frac{\mathcal{R}_n}{2}e^{\left(\frac{4}{n}+2\right)\alpha\phi}\right)\,.\label{appendix_final_EH_action}
\end{align}
Finally, one can choose $\alpha$ such that the scalar field is canonically normalized:
\begin{align}
      \alpha^2=\frac{n}{2(n+2)}\,.\label{appendix_alpha_eq}
\end{align}
\paragraph{Gibbons-Hawking-York boundary term:}
We continue with the reduction of the Gibbons-Hawking-York boundary term from $d$ to four dimensions, assuming that the boundary is a surface defined by $r=const.$
We can directly calculate
\begin{align}
    S_{GHY}=-\int_\Sigma \sqrt{h}\mathcal{K}_d=-\int_\Sigma \sqrt{g_n}e^{3\alpha\phi}e^{n\beta\phi}f^3\mathcal{K}_d\,,\label{appendix_Kd}
\end{align}
with $h$ the induced metric on the $(d-1)$-dimensional boundary.
To calculate the $d$-dimensional extrinsic curvature $\mathcal{K}_d$, we note that the outward pointing normal vector is $n=e^{-\alpha\phi}\partial_r$.
We further use $3\alpha+n\beta=\alpha$ from \eqref{appendix_alpha_beta_rel}, resulting in
\begin{align}    
\mathcal{K}_d=\frac{1}{e^{(4\alpha+n\beta)\phi}f^3}\partial_r \left(e^{(3\alpha+n\beta)\phi}f^3\right)= e^{-\alpha\phi}\left(\alpha\phi'+3\frac{f'}{f}\right)\,,
\end{align}
and hence
\begin{align}
    -\int_\Sigma \sqrt{h}\mathcal{K}_d=-\int f^3\alpha\phi'-\int f^3\mathcal{K}_4\,.\label{appendix_GHY_redced}
\end{align}
One now observes that the term $\alpha\square_4\phi$ in \eqref{appendix_final_EH_action} is a total derivative and contributes to the action as 
\begin{align}
    \left.\alpha f^3\phi'\right|_{r=0}^{r=\infty}\,.\label{appendix_box_contribution}
\end{align}
Hence, for Minkowski compactifications, the first term on the r.h.s. of~\eqref{appendix_GHY_redced} precisely cancels the contribution from the $r=\infty$ term of \eqref{appendix_box_contribution}.

Since, from the 4d perspective, the bubble wall (for bubbles of anything) is another boundary of spacetime, it appears logical to also include a corresponding GHY term in the four-dimensional action. In addition, the bubble wall has an effective 4d tension, such that we supplement our total action with a term 
\begin{align}
-\int_{\partial \Sigma}\sqrt{g_4}\,({\cal K}_4-T_4)=
    -2\pi^3 f^3 \left(\mathcal{K}_4-T_4\right)\label{appendix_new_boundary}
\end{align}
on the boundary corresponding to the bubble wall.
At the same time, we remove the term $\alpha\square_4\phi$ from the bulk action since we take its remaining effect at the bubble wall from \eqref{appendix_box_contribution} into account in the new boundary term \eqref{appendix_new_boundary}.

\section{Bubble of nothing decays in general dimensions}\label{appendix_sect_general_dimension}
In this section, we collect the relevant equations to analyze bubble of nothing decays in general dimensions.

For general $d$, Equations \eqref{phi_prime_solution} and \eqref{phi_f} are still valid, just the values of $\alpha$ and $\beta$ change according to \eqref{alpha_beta_equations}.
Demanding that the geometry ends when the extra dimensions reach critical size $\eta$ results in the defining equation
\begin{align}
    2\pi\eta=2\pi R(f_\eta) = e^{\beta\phi(f_\eta)}\,\label{appendix_eta_def}
\end{align}
for $f_{\eta}$.
Equation \eqref{appendix_eta_def} simplifies to
\begin{align}
    \frac{2C}{\sqrt{6}f_\eta^2}=\left(\frac{R_{KK}}{\eta}\right)^{\frac{2}{\sqrt{6}\beta}}-\left(\frac{R_{KK}}{\eta}\right)^{-\frac{2}{\sqrt{6}\beta}}\,.\label{appendix_C_f2}
\end{align}
Unless $\eta$ is extremely close to $R_{KK}$, we can neglect either the first or the second term in the equation above.
In the main text, $\eta/R_{KK}<1$ was assumed to improve readability of formulae. 
In the following, we will present the general equations.

Introducing the (analogue of) the deficit angle $\theta$ as in \eqref{theta_def_CY} 
\begin{align}
    1-\frac{\theta}{2\pi}=\left.\frac{dR}{dr}e^{-\alpha\phi}\right|_{r=0}=\frac{\beta}{2\pi}\frac{C}{f_\eta^3}(2\pi\eta)^{1+\frac{n}{2}}
\end{align}
and solving for $C$ by using \eqref{appendix_C_f2} results in
\begin{align}
    C=\frac{\beta^2}{(2\pi)^2}\frac{3\sqrt{6}}{4\left(1-\frac{\theta}{2\pi}\right)^2}(2\pi\eta)^{n+2}\left[\left(\frac{R_{KK}}{\eta}\right)^{\frac{2}{\sqrt{6}\beta}}-\left(\frac{R_{KK}}{\eta}\right)^{-\frac{2}{\sqrt{6}\beta}}\right]^3\,.
\end{align}
The extrinsic curvature in four dimensions can now be calculated to be
\begin{align}
    \mathcal{K}_4=-\sqrt{6}\frac{2\pi\left|1-\frac{\theta}{2\pi}\right|}{\beta}(2\pi\eta)^{-1-\frac{n}{2}}\sqrt{\left[\left(\frac{R_{KK}}{\eta}\right)^{\frac{2}{\sqrt{6}\beta}}-\left(\frac{R_{KK}}{\eta}\right)^{-\frac{2}{\sqrt{6}\beta}}\right]^{-2}+\frac{1}{4}}\,.\label{appendix_K4}
\end{align}
Interestingly, $\mathcal{K}_4$ does not depend on $R_{KK}$, unless $\eta\approx R_{KK}$. 
This provides evidence to the claim that bubbles of anything are local quantum effects.

The critical bubble radius $\rho_0$ is computed to be
\begin{align}
    \rho_0=\sqrt{\frac{3}{2}}\frac{\beta}{\left|1-\frac{\theta}{2\pi}\right|}\eta\left|\left(\frac{R_{KK}}{\eta}\right)^{\frac{2}{\sqrt{6}\beta}}-\left(\frac{R_{KK}}{\eta}\right)^{-\frac{2}{\sqrt{6}\beta}}\right|\,.
\end{align}
Finally, the decay exponent is given by
\begin{align}
\begin{split}
    B&=-2\pi^2f_{\eta}^3\frac{\mathcal{K}_4}{3}\\
    &=\frac{3}{4}\frac{\beta^2}{\left(1-\frac{\theta}{2\pi}\right)^2}(2\pi\eta)^{n+2}\sqrt{\left[\left(\frac{R_{KK}}{\eta}\right)^{\frac{2}{\sqrt{6}\beta}}-\left(\frac{R_{KK}}{\eta}\right)^{-\frac{2}{\sqrt{6}\beta}}\right]^{4}+\frac{1}{4}\left[\left(\frac{R_{KK}}{\eta}\right)^{\frac{2}{\sqrt{6}\beta}}-\left(\frac{R_{KK}}{\eta}\right)^{-\frac{2}{\sqrt{6}\beta}}\right]^{6}}\,.
    \end{split}\label{appendix_B_general}
\end{align}
Inspecting the profile of the radius $\rho(f)$ of the transversal $S^3$
\begin{align}
    \rho(f)=R_{KK}^{-\frac{n}{2}}\left(\frac{C}{\sqrt{6}f^2}+\sqrt{\left(\frac{C}{\sqrt{6}f^2}\right)^2+1}\right)^{-\alpha\frac{\sqrt{6}}{2}}f
\end{align}
shows that for large $f$
\begin{align}
    \rho(f)\approx R_{KK}^{-\frac{n}{2}}f\,,
\end{align}
and hence $\rho$ grows with growing $f$.
On the other hand, if $f_\eta^2\ll C$, there will be the regime where
\begin{align}
    \rho(f)\approx 
    \begin{cases}
        R_{KK}^{-\frac{n}{2}} \left(\frac{2C}{\sqrt{6}}\right)^{-\alpha\sqrt{6}/2}f^{1+\alpha\sqrt{6}} & C>0\,,\\
        R_{KK}^{-\frac{n}{2}}\left(\frac{\sqrt{6}}{2C}\right)^{-\alpha\sqrt{6}/2}f^{1-\alpha\sqrt{6}} & C < 0\,.
    \end{cases}
\end{align}
It now depends on the value of $\alpha$ whether $\rho$ grows or shrinks with growing $f$.
We first note that $\alpha<0$ for all values of $d$, such that $\rho$ grows with growing $f$ if $C<0$.
Let us now analyze the case $C>0$:
If, $d=5$, $\alpha=-\frac{1}{\sqrt{6}}$ and hence $\rho$ stays approximately constant.
Since $\alpha < -1/\sqrt{6}$ for $d>5$, one observes that $\rho$ shrinks with growing $f$ near the bubble wall if $C>0$.
Such a peculiar feature has already been observed in \cite{Blanco-Pillado:2011fcm}.

\section{Supersymmetry and the ETW brane}\label{appendix_sect_SUSY_O3_O5}
In this Appendix, we show that a spacetime filling O3/D3 system together with a O5/D5 domain wall wrapped on a special Lagrangian three-cycle of a Calabi-Yau manifold preserves two supercharges. We mostly follow the ideas of \cite{Martucci:2005ht,Tomasiello:2022dwe}.
We further argue that this situation also holds for O5/D5, O7/D7 models.
Since Op planes preserve the same supercharges as Dp branes, we can equivalently think of D-brane systems only.
The two SUSY generators of type II theories will be denoted by $\epsilon,\tilde\epsilon$. 
Then, the condition that a brane preserves supersymmetry is given by
\begin{align}
    \tilde\epsilon=\Gamma_\parallel^{Dp}\epsilon\,,\qquad\Gamma_\parallel^{Dp} = \frac{1}{\sqrt{g|_{Dp}}}\prod_i\Gamma_i\,,
\end{align}
where the product is over all gamma matrices in directions parallel to the Dp brane.
After compactifying on a Calabi-Yau, the SUSY generators for type IIB string theory can be parametrized as
\begin{align}
    \epsilon &= \zeta_+\otimes \eta_+ + \zeta_-\otimes \eta_-\,,\\
    \tilde\epsilon&= \tilde\zeta_+\otimes \eta_+ + \tilde\zeta_-\otimes \eta_-\,.
\end{align}
Here, $\zeta_+,\tilde\zeta_+$ are independent four-dimensional spinors of positive chiraliy and $\eta_+$ is the unique six-dimensional covariantly constant positive-chirality spinor of the Calabi-Yau. 
Further, $\zeta_-=(\zeta_+)^c,\tilde\zeta_-=(\tilde\zeta_+)^c$, $\eta_-=(\eta_+)^c$.
The eight real degrees of freedom of $\zeta_+,\tilde\zeta_+$ lead to the familiar $\mathcal{N}=2$ theory in 4d.
Using similar arguments as in \cite{Tomasiello:2022dwe} chapter 9.2 and adopting the same conventions, one finds that the condition
\begin{align}
    \tilde\epsilon=\Gamma_\parallel^{D3}\epsilon\,,
\end{align}
where the D3 brane is spacetime filling, results in the relation
\begin{align}
    \tilde\zeta_+=i\zeta_+\,.\label{appendix_D3_condition}
\end{align}
As a result, the preserved supercharges are reduced from eight to four.
The second condition 
\begin{align}
    \tilde\epsilon=\Gamma_\parallel^{D5}\epsilon\,,
\end{align}
with the D5 brane being extended in the $0,1,2$ directions of the non-compact space and wrapping a three-cycle $S_3$ of the CY, implies
\begin{align}
    \zeta_-=\beta\gamma_3\zeta_+\,,\quad \eta_-=\beta^{-1}\gamma_\parallel\eta_+\,,\label{appendix_D5_condition}
\end{align}
with $\beta$ a constant to be determined.
Here, following \cite{Tomasiello:2022dwe} chapter 9.2 and 5.6, $\gamma_3$ is to be seen as a 4d gamma matrix and $\gamma_\parallel$ is the product of gamma matrices along the directions of the three-cycle $S_3$.
Since $\gamma_\parallel$ is unitary, the second condition in \eqref{appendix_D5_condition} determines $|\beta|=1$.
Hence, $\beta=e^{i\phi}$.
It is now straightforward to see that the first condition of \eqref{appendix_D5_condition} always has a solution. 
For example, in the chiral basis, we have the following representation of $\gamma_3$ and a general spinor $\zeta_+,\zeta_-$
\begin{align}
    \gamma_3=i\begin{pmatrix}
        0 & -\sigma^3\\
        \sigma^3 & 0
    \end{pmatrix}\,,\qquad\zeta_+=\begin{pmatrix}
        a\\
        b\\
        0\\
        0
    \end{pmatrix}\,,\qquad\zeta_-\begin{pmatrix}
        0\\
        0\\
        -\overline{b}\\
        \overline{a}
    \end{pmatrix}\,,\qquad a,b\in \mathbb C\,.
\end{align}
The first condition of \eqref{appendix_D5_condition} then becomes
\begin{align}
    \begin{pmatrix}
        0\\
        0\\
        -\overline{b}\\
        \overline{a}
    \end{pmatrix}=e^{i\phi}\begin{pmatrix}
        0\\
        0\\
        ia\\
        -ib
    \end{pmatrix}\,.
\end{align}
We see that this system can be solved for any $\phi$ by choosing $a=ie^{-i\phi}\overline{b}$.
This condition further reduces the number of preserved supercharges from four to two, resulting in $\mathcal{N}=1$ supersymmetry in 3d.
Finally, we need to check whether the second condition of \eqref{appendix_D5_condition} is fulfilled. 
Following the logic of \cite{Tomasiello:2022dwe} chapter 5.6, 9.2 and adopting it to our case of interest, one finds that $\eta_-=e^{-i\phi}\gamma_\parallel\eta_+$ is obeyed when the cycle $S_3$ is calibrated with respect to the form $Re(e^{i\theta}\Omega)$
\begin{align}
    Re(e^{i\phi}\Omega)|_{S_3}=\text{vol}_{S_3}\,.\label{sLag_condition}
\end{align}
The condition \eqref{sLag_condition} precisely states that the three cycle $S_3$ needs to be special Lagragian. 
We conclude that our D3/D5 system preserves $\mathcal{N}=1$ SUSY in 3d if the D5 brane is wrapped on a special Lagrangian cycle.
Since D7 branes wrapped on holomorphic four-cycles of the CY preserve the same supercharges as D3 branes, our results straightforwardly generalize to such D7/D5 models.
This can further be seen by again using techniques from \cite{Tomasiello:2022dwe} chapter 9 and by noting that the condition
\begin{align}
    \tilde\epsilon = \Gamma_\parallel^{D7}\epsilon\,,
\end{align}
with the D7 wrapped on a holomorphic cycle, also leads to \eqref{appendix_D3_condition}.
Then, from the arguments below \eqref{appendix_D3_condition}, it is clear that two supercharges are preserved if the $D5$ brane is wrapped on a special Lagrangian cycle of the CY.

\section{AdS bubble transitions in the limit of small AdS length}\label{appendix_sect_AdS_negative_tension}

In this Appendix we discuss whether and how tunneling to/from a deep AdS may be related to tunneling to/from `nothing'. This analogy has been suggested and analysed in\cite{Brown:2011gt}, but we disagree with a key detail which is decisive in our context.

We start by considering thin-wall tunneling instantons for the decay of Minkowski space to AdS as a possible model for bubbles of nothing.
We denote the bubble radius by $R$, the AdS length by $\ell_{AdS}$, the matter tension by $T$ and we work in 4d Planck units.
The tunneling exponent prior to radius extremization is given by (see e.g. \cite{Eckerle:2020opg})
\begin{align}
    B_{\text{Mink. $\to$ AdS}}=6\pi^2R^2+2\pi^2TR^3-4\pi^2\ell_{AdS}^2\left(\left(1+\frac{R^2}{\ell_{AdS}^2}\right)^{3/2}-1\right)\,.\label{appendix_AdS_tunneling_action}
\end{align}
The instanton corresponds to a patch of AdS of radius $R$ glued into Minkowski space.
The first term on the r.h.s.~of \eqref{appendix_AdS_tunneling_action} represents the extrinsic curvature contribution at the bubble wall on the Minkowski side, the second term is the effect of the wall tension, and the third term combines both the AdS bulk and the AdS-side extrinsic curvature contributions.  The extrinsic curvature term at infinity is cancelled by the Minkowski vacuum contribution.  In the limit $R\gg \ell_{AdS}$ the last term in~\eqref{appendix_AdS_tunneling_action} reduces to $2\pi^2(-2/\ell_{AdS})R^3$. As a result, the total effect of the AdS side is to correct the brane tension by $-2/\ell_{AdS}$. Thus, at $R\gg \ell_{AdS}$ one has
\begin{align}
B_{\text{Mink. $\to$ AdS}} =6\pi^2R^2+2\pi^2T_{eff}R^3\,,\label{appendix_AdS_tunneling_action_eff}
\end{align}
with $T_{eff}=T-2/\ell_{AdS}$. One immediately sees that this is the exponent for a bubble-of-nothing tunneling process in flat space.

The action \eqref{appendix_AdS_tunneling_action} has an extremum only for $T_{eff}<0$, in which case the critical bubble radius reads
\begin{align}
    R_{\text{Mink. $\to$ AdS}}=\frac{4T}{\frac{4}{\ell_{AdS}^2}-T^2}=-\frac{4\left(\frac{T_{eff}\ell_{AdS}}{2}+1\right)}{\frac{\ell_{AdS}T_{eff}^2}{2}+2T_{eff}}\,.
\end{align}
In the limit $\ell_{AdS}\to 0$, one then obtains
\begin{align}
    R_{\text{Mink. $\to$ AdS}}=-\frac{2}{T_{eff}}\,.
\end{align}
This could have equivalently been found by extremizing \eqref{appendix_AdS_tunneling_action_eff}.
We see that tunneling becomes possible if the effective tension is negative. This effective tension may be thought of as a physical quantity by taking the renormalization-type limit $\ell_{AdS}\to 0$, $T\to \infty$ while keeping $T_{eff}$ finite.

For the bubble of something, the corresponding AdS $\to$ Minkowski tunneling exponent takes the form \begin{align}
    B_{\text{AdS $\to$ Mink.}}=-6\pi^2R^2+2\pi^2TR^3+4\pi^2\ell_{AdS}^2\left(\left(1+\frac{R^2}{\ell_{AdS}^2}\right)^{3/2}-1\right)\,.\label{appendix_AdS_tunneling_action_BoS}
\end{align}
The instanton geometry is a patch of Minkowski space of radius $R$ glued into AdS, which in contrast to \cite{Brown:2011gt} we take to be infinite. Extremizing the expression in~\eqref{appendix_AdS_tunneling_action_BoS} gives
\begin{align}
    R_{\text{AdS $\to$ Mink.}}=\frac{4T}{T^2-\frac{4}{\ell_{AdS}^2}}\label{appendix_R_crit_bos}
\end{align}
for $T<0$ and no extremum otherwise. Moreover, demanding a positive bubble radius one finds the additional requirement $T_{eff}\equiv T+2/\ell_{AdS}>0$. 
In the limit $\ell_{AdS}\to 0$, the tunneling exponent \eqref{appendix_AdS_tunneling_action_BoS} becomes
\begin{align}
    B_{\text{AdS $\to$ Mink.}}    =-6\pi^2R^2+2\pi^2T_{eff}R^3\,.\label{appendix_AdS_tunneling_action_eff_bos}
\end{align}
It is extremized by
\begin{align}
    R_{\text{AdS $\to$ Mink.}}=\frac{2}{T_{eff}}\,,
\end{align}
which is also the $\ell_{AdS}\to 0$ limit of \eqref{appendix_R_crit_bos} if $T_{eff}$ is kept finite. We see that the limit $\ell_{AdS}\to 0$ can only be sensibly taken when the `bare tension' $T$ is negative and diverges. While negative-tension ETW-branes are known to exist, it is not clear to us whether the negative tension brane between AdS and Minkowski required above can be realized in string theory or even makes sense more generally, in an EFT. 

Our key conclusion for the present paper is that bubbles of something are not ruled out since, if negative tension branes are forbidden, then up-tunneling from AdS is not a good model for bubbles of something. If, on the contrary, negative-tension branes are allowed, then up-tunneling from AdS reproduces the simpler bubble-of-something results.

\bibliographystyle{utphys}
\bibliography{References_BON}
\end{document}